\documentclass[preprint2]{emulateapj}
\usepackage{graphicx}
\usepackage{natbib}

\def\cs{$c_{\rm s}$ }
\def\solmas{$\mathrm{M_\odot}$\,}
\def\solmasp{$\mathrm{M_\odot}$}

\def\apjl{{\em ApJ}}
\def\apjs{{\em ApJS}}
\def\aap{{\em A\&A}}

\def\simless{\mathbin{\lower 3pt\hbox
   {$\rlap{\raise 5pt\hbox{$\char'074$}}\mathchar"7218$}}}
\def\simgreat{\mathbin{\lower 3pt\hbox
   {$\rlap{\raise 5pt\hbox{$\char'076$}}\mathchar"7218$}}}

\newcommand{\hd}{\rm{HD}}
\newcommand{\mD}{\rm{D}}
\newcommand{\mH}{{\rm H}}
\newcommand{\Hp}{{\rm H}^{+}}
\newcommand{\Dp}{{\rm D}^{+}}
\newcommand{\mHt}{{\rm H_{2}}}
\newcommand{\me}{{\rm e^{-}}}
\newcommand{\Hm}{{\rm H}^{-}}
\newcommand{\He}{{\rm He}}
\newcommand{\Hep}{{\rm He^{+}}}
\newcommand{\Hepp}{{\rm He^{++}}}
\newcommand{\mHtp}{{\rm H_{2}^{+}}}

\def\simless{\mathbin{\lower 3pt\hbox
   {$\rlap{\raise 5pt\hbox{$\char'074$}}\mathchar"7218$}}}
\def\simgreat{\mathbin{\lower 3pt\hbox
   {$\rlap{\raise 5pt\hbox{$\char'076$}}\mathchar"7218$}}}

\begin{document}

\title{Gravitational fragmentation in turbulent primordial gas and the initial mass function of Population III stars}

\author
{Paul C.\ Clark$^1$, Simon C.O. Glover$^1$,  Ralf S.\ Klessen$^{1, 2}$ \& Volker Bromm$^3$}

\affil{
$^{1}$ Institut f\"ur Theoretische Astrophysik, Zentrum f\"ur Astronomie der Universit\"at Heidelberg, Albert-Ueberle-Str.\ 2, 69120 Heidelberg, Germany.
\break email: pcc@ita.uni-heidelberg.de, sglover@ita.uni-heidelberg.de, rklessen@ita.uni-heidelberg.de \\
$^{2}$ Kavli Institute for Particle Astrophysics and Cosmology, Stanford University, Menlo Park, CA 94025, USA. \\
$^{3}$ The University of Texas, Department of Astronomy and Texas Cosmology Center, 2511 Speedway, RLM 15.306, Austin, TX 78712, USA.
\break email: vbromm@astro.as.utexas.edu
}

\begin{abstract}

We report results from numerical simulations of star formation in the early universe that focus on the dynamical behavior of metal-free gas under different initial and environmental conditions. In particular we investigate the role of turbulence, which is thought to ubiquitously accompany the collapse of high-redshift halos. We distinguish between two main cases: the birth of Population III.1 stars -- those which form in the pristine halos unaffected by prior star formation -- and the formation of Population III.2 stars -- those forming in halos where the gas has an increased ionization fraction.  We find that turbulent primordial gas is highly susceptible to fragmentation in both cases, even for turbulence in the subsonic regime, i.e.\ for rms velocity dispersions as low as 20 \% of the sound speed. Fragmentation is more vigorous and more widespread in pristine halos compared to pre-ionized ones. If such levels of turbulent motions were indeed present in star-forming minihalos, Pop III.1 stars would be on average of somewhat lower mass, and form in larger groups, than Pop III.2 stars. We find that fragment masses cover over two orders of magnitude, suggesting that the Population~III initial mass function may have been much broader than previously thought. This prompts the need for a large, high-resolution study of the formation of dark matter minihalos that is capable of resolving the turbulent flows in the gas at the moment when the baryons become self-gravitating. This would help to determine the applicability of our results to primordial star formation
\end{abstract}

\maketitle

\section{Introduction}
\label{sec:intro}

Over the course of the last decade, work by a number of groups has led
to the development of a widely accepted picture for the formation of the
first stars, the so-called Population III (or Pop.\ III) stars. In this
picture, the very first stars (Population III.1 in the nomenclature 
of \citealt{tm08} and \citealt{omha2008}) form within small dark matter halos that have total 
masses $M \sim 10^{6} \: {\rm M_{\odot}}$, virial temperatures of around
a thousand K, and that are assembled at redshifts $z \sim 25$--30 or 
above \citep{bl04,g05,byhm09}. Gas falling into one of these small dark
matter halos is shock-heated to a temperature close to the virial temperature
of the halo, and thereafter cools via H$_{2}$ rotational and vibrational
line emission. The H$_{2}$ that enables the gas to cool is primarily 
formed via the gas-phase reactions
\begin{eqnarray}
{\rm H} + {\rm e^{-}} & \rightarrow & {\rm H^{-}} + \gamma, \\
{\rm H^{-}} + {\rm H} & \rightarrow & {\rm H_{2}} + {\rm e^{-}},
\end{eqnarray}
where the required free electrons are those that remain in the gas
after cosmological recombination at $z \sim 1100$. The low abundance
of these electrons, plus the fact that the gas starts to recombine
further once it collapses into the dark matter halo, limits the
amount of H$_{2}$ that can form in this manner; the typical fractional
abundances of H$_{2}$ that result are a few times $10^{-3}$. The 
limited H$_{2}$ abundance, the low critical density of the H$_{2}$ 
molecule, and the large energy separation of its lowest accessible
rotational levels combine to significantly limit the extent to which 
the gas can cool. The minimum temperature reached depends upon the
dynamical history of the collapse, but is typically around $200 \: {\rm K}$.
This minimum temperature is reached at a density of around $10^{4}
\: {\rm cm^{-3}}$, comparable to the critical density of H$_{2}$
(the density at which its level populations reach their local 
thermodynamical equilibrium (LTE) values), and at higher densities,
the gas begins to reheat. Gas falling into the dark matter halo 
therefore tends to fragment once it reaches this temperature and
density, and the resulting fragments have masses of the order of
the local Jeans mass, $M_{\rm J} \sim 1000 \: {\rm M_{\odot}}$.\footnote{Note 
that the term `fragmentation' is often used in the literature even in the case 
where only one object forms at the center of the clump, even though, strictly
speaking, the term refers to a system separating into several distinct parts
with separate evolutionary paths.}
Following this period of fragmentation, the gas in the fragments undergoes
a further period of gravitational collapse, eventually reaching a density of
$n \sim 10^{8}$--$10^{10} \: {\rm cm^{-3}}$, at which point three-body H$_{2}$
formation becomes effective. This rapidly converts almost all of the available
atomic hydrogen into H$_{2}$, but the significant energy release that accompanies
this process heats the gas, which typically attains a temperature of 1000--2000~K
during this phase of the collapse. 

Continued collapse next leads to the gas becoming optically thick in the main
H$_{2}$ cooling lines (at $n \sim 10^{10} \: {\rm cm^{-3}}$; \citealt{ra04}, 
\citealt{yoha06}), the onset of collision-induced emission cooling (at $n
\sim 10^{14} \: {\rm cm^{-3}}$; \citealt{ra04}), and finally to the gas becoming
optically thick in the continuum as well as in the H$_{2}$ lines ($n \sim 10^{16}
\: {\rm cm^{-3}}$; \citealt{yoh08}). H$_{2}$ dissociation cooling allows for
further collapse, but once all of the H$_{2}$ is gone, the collapse becomes fully
adiabatic. Detailed simulations of the collapse, using adaptive mesh
refinement (AMR; see e.g.\ \citealt{abn02, on07}) or smoothed particle 
hydrodynamics (SPH; see e.g.\ \citealt{brlb04,yoha06,yoh08}) 
show little or no evidence for fragmentation between
$n \sim 10^{4} \: {\rm cm^{-3}}$ and the onset of this adiabatic evolution, and
it is natural to associate the latter with the formation of a primordial protostar.
The amount of gas incorporated into the protostar at this point is small, 
approximately $0.01 \: {\rm M_{\odot}}$, but it is surrounded by a massive, dense
envelope and hence begins to accrete rapidly. If the envelope does not fragment,
and if accretion onto the protostar is unhindered by radiative feedback, then
the final mass of the star can be very large. For example, \citet{on07} estimate
that its mass could lie anywhere in the range $M \sim 20$--$2000 \: {\rm M_{\odot}}$,
depending on the environment and dynamical history of the gas. The degree
to which feedback will suppress accretion remains uncertain, but the most 
effective potential feedback mechanisms, Lyman-$\alpha$ scattering and
photoionization \citep{mt08} become important only once the mass of the star
exceeds 20~M$_{\odot}$.  The expectation is therefore that all Population III.1 stars were
massive, with masses typically of the order of  $100 \: {\rm M_{\odot}}$ or more, 
thereby also explaining why none of these stars appear to have survived until the 
present day.  If a Population III.1 star of 
this mass does form, then it will rapidly dissociate H$_{2}$ throughout the halo, 
thereby suppressing further star formation \citep{on99,gb01}. This fact, 
together with the lack of fragmentation seen in the simulations, is often taken to
imply that only a single Population III.1 star will form in each dark matter halo.

This widely accepted picture for Population III star formation also provides for
a second mode of Pop.\ III star formation. This occurs in metal-free gas which 
has been ionized by radiation from a previous generation of Population III
stars. Following the death of these stars, the gas recombines, and the elevated
fractional ionization in the recombining gas allows more H$_{2}$ to form. The
enhanced H$_{2}$ fraction enables the gas to cool to a lower temperature,
which in turn increases the effectiveness of cooling by the singly-deuterated 
hydrogen molecule, HD. In the standard Pop.\ III.1 scenario, HD cooling is
of limited importance, but in this second scenario, termed Population III.2
star formation by \citet{tm08}, it becomes dominant,  cooling the gas down 
to the temperature of the cosmic microwave background
\citep{no05,jb06,yoh07}. The lower gas temperature, plus the higher critical density 
of HD ($n_{\rm crit} \sim 10^{6} \: {\rm cm^{-3}}$, compared with $n_{\rm crit} \sim 10^{4} \: 
{\rm cm^{-3}}$ for H$_{2}$), means that fragmentation occurs somewhat later,
and produces significantly lower mass fragments, with characteristic masses
$M \sim 100 \: {\rm M_{\odot}}$. Subsequently, the evolution of these fragments
is believed to proceed in the same fashion as described above, resulting 
ultimately in the formation of a primordial protostar of similar mass. However, 
since this protostar is embedded in a less massive envelope, with a lower
accretion rate, the final stellar mass is thought to be an order of magnitude
or so smaller, $M \sim 10 \: {\rm M_{\odot}}$. Nevertheless, this still corresponds to what we
would call, by Galactic standards, a massive star, and moreover one which
is more than capable of dissociating H$_{2}$ throughout a large volume of
the halo, thereby suppressing further star formation. 

In the past few years, it has become possible to partially test this picture
using numerical simulations that begin with the proper cosmological initial
conditions and follow the collapse of the gas all the way to protostellar
densities. These simulations confirm that fragmentation during the initial
collapse phase is ineffective: in general, only a single protostar forms,
although in roughly 20\% of cases, the gas fragments into two clumps, 
forming a wide binary (\citealt{tao09}; M.~Turk, private communication). 
However, most of these studies were prevented by technical limitations 
from following the evolution of the infalling gas after the formation of the first 
protostar. The technical problem involves the hydrodynamical timestep: as the 
collapse is followed down to protostellar densities, this becomes extremely short
($\Delta t \sim 10^{-3} \: {\rm yr}$), making it computationally infeasible 
to follow the evolution of the surrounding dense gas over any significant 
time period.

In studies of contemporary star formation, a similar problem is avoided
by the use of sink particles \citep{bbp95}. When gravitationally bound regions 
of gas collapse below the scale on which they can be spatially resolved, they
are replaced in the simulation by a sink particle of the same mass. This
can accrete gas from its surroundings, but otherwise interacts only via 
gravity. Recent simulations of Population III star formation that have used 
sink particles to allow the evolution of the gas to be followed beyond the 
formation of the first protostar consistently find evidence for more extensive 
fragmentation than is assumed in the conventional picture of Population III star 
formation (\citealt{cgk08,sgb10}; Clark et~al., in prep.). If fragmentation is as 
common as these results suggest, and most Population III stars form in multiple 
systems rather than as single stars, then this has profound implications for 
the final masses of the stars, their production of ionizing photons and metals,
the rate of high redshift gamma ray bursts, and many other issues.

It is therefore important to better understand the physical basis of
fragmentation in these systems. To do this, we would ideally like
to have a large sample of simulations, as basing our arguments on
only one or two realizations of Population III star formation leaves one
open to the possibility that these  realizations may not be typical. 
Unfortunately, simulations starting from cosmological initial conditions
and following collapse all the way to protostellar densities are computationally
costly, making it difficult to explore a large region of parameter space.
In these simulations, it is also difficult to be certain about which aspects
of the included physics are the most important for driving fragmentation.
Simulations that start from simpler initial conditions and that focus on
exploring the importance of a single free parameter therefore play an
important role, which is complementary to that of fully cosmological
simulations. A good example is the recent work by Machida and 
collaborators \citep{mach08,mach09} which examined the influence
of the initial rotational energy of the gas, and found that zero metallicity
clouds with sufficient initial rotational energy could fragment into tight
binaries. Another example, and one which explores a much lower density regime, is the
study by \citet{jap09a, jap09b}, who showed that the fragmentation behavior
depends sensitively on the adopted density profile of the primordial halo. 

In this paper, we perform a similar study into the effects of the initial
turbulent energy of the gas. Our motivation comes from the fact that 
recent high-resolution cosmological simulations have shown that the 
self-gravitating regions in which Pop.\ III star formation occurs contain 
significant turbulent motions \citep[see e.g.][]{2008MNRAS.387.1021G}.  
From studies of present-day star formation, we know that such turbulent motions 
can lead to the fragmentation of initially marginally unstable gas clouds 
\citep{Klessen01,2004A&A...414..633G,2004A&A...423..169G,mk04,2009A&A...495..201A}, 
and so it is instructive to investigate how such conditions could 
affect the `standard' picture of primordial star formation. In a Pop.\ III analogue of the 
studies on present-day star formation, we will look at how subsonic turbulence affects the
collapse of marginally super-critical Bonnor-Ebert spheres, in an attempt to quantify the effects
of turbulently induced fragmentation on the mass function of Pop.\ III stars. Central to 
this study will be the use of the sink particles to allow us to capture the evolution of the
gas cloud beyond the collapse of the first region.

The paper is laid out in the following manner. Our modifications to the cosmological SPH code Gadget 2 \citep{springel05} are outlined in \S \ref{sec:gadget2} (with further details of the chemical networks and heating and cooling described in the Appendix). The initial conditions for this numerical experiment are described in \S \ref{sec:ics}, including the definitions of Pop.\ III.1 and III.2 that are used in our study. The details of the fragmentation process seen in the simulations are discussed in \S \ref{sec:frag} and the long-term chemical and thermodynamical evolution of the infalling envelope are described in \S \ref{sec:evol}. We discuss the implications of this study for the initial mass
function (IMF) of primordial stars in \S \ref{sec:chat}, and summarize the main points of this paper in \S \ref{sec:sum}.

\section{Numerical method}
\label{sec:gadget2}
We model the evolution of the gas in our simulations using a modified version
of the Gadget 2 smoothed particle hydrodynamics code \citep{springel05}. 
We have modified the publicly available code in several respects. First, we have
added a sink particle implementation, based on the prescription in \citet{bbp95},
to allow us to follow the evolution of the gas beyond the point at which the first
protostar forms. Our particular implementation is derived from the one first described in
\citet{jap05}. We briefly describe the ideas behind the algorithm here. The actual numerical values used for the parameters discussed here are given later, in \S~\ref{sec:ics}.

Sink particles are not really added to the code in the sense that a new
particle is introduced, but instead a normal SPH particle is turned into a sink
particle once certain criteria have been met. The particle undergoes a series
of tests once it has reached a threshold density. The first is to see whether the
candidate SPH particle is sufficiently far away from any other sinks, measured in terms of the
sink particle's accretion radius, $r_{\rm acc}$. We adopt a
conservative value of $2 r_{\rm acc}$. Next, we check to see whether the smoothing
length of the particle is less than the `accretion radius' of the sink
particle that it will become. This ensures that when the sink particle forms it
can instantly accrete at least $\sim 50$ neighboring particles (we adopt 50 neighbors in these simulations). The third
test is to make sure that the candidate sink particle and its neighbors are on
the same integration time-step.

Once these three preliminary criteria are met, the dynamical state of the possible sink
particle and its neighbors are assessed to ensure that the particles are
indeed undergoing gravitational collapse and are not about to re-expand from
their dense state. This takes the form of a further four tests. Firstly, we require that

\begin{equation}
\alpha \le \frac{1}{2}
\end{equation}

\noindent where $\alpha$ is the ratio of the thermal energy to the magnitude of
the gravitational energy of the particles. Secondly, we ensure that

\begin{equation}
\alpha + \beta \le 1
\end{equation}

\noindent where $\beta$ is the ratio of rotational energy to the magnitude of
the gravitational energy. The third condition is that the total energy of the
particles must be negative (which actually renders the above checks redundant, but can help to improve the computational efficiency of the code). Finally, the forth test is that the divergence of the accelerations must be less than zero. This final check ensures that the
group of particles is not in the process of being tidally disrupted or
bouncing. If all of these tests are passed, the particle becomes a sink and the
mass and linear momentum of the neighbors are added to the sink
particle.

As the simulation progresses, the sink particles are then allowed to accrete other gas particles that fall within the accretion radius. As in the Bate et al. (1995) prescription, several tests must be passed before any SPH particles can be accreted by the sink. First, it must obviously be bound and moving toward the candidate sink. Second, in the case where there is more than one sink present, it must be more bound to the candidate sink than any other sink in the simulation. Finally, the SPH and sink particles need to be on the same integration time-step (to ensure temporal momentum conservation). Once these conditions are passed, the mass and linear momentum of the SPH particle are added to the sink particle. Note that accretion onto existing sink particles is done {\em before} any new candidate sinks are considered.

In addition to the sink particles, we have also implemented an external pressure term \citep[e.g.][]{benz90}, that enables us to model a constant pressure boundary, as opposed to the vacuum or periodic boundary conditions that are the only choices available in the standard version of Gadget. We modify the standard gas pressure contribution to the Gadget 2 momentum equation, 

\begin{equation}
\frac{d v_{i}}{d t} = - \sum_{j} m_{j} \left[
f_{i}\frac{P_{i}}{\rho_{i}^{2}} \nabla_{i} W_{ij}(h_{i})
+ f_{j}\frac{P_{j}}{\rho_{j}^{2}}\nabla_{i} W_{ij}(h_{j}) \right] , 
\end{equation}

\noindent by replacing $P_{i}$ and $P_{j}$ with $P_{i} -  P_{\rm ext}$ and $P_{j} -  P_{\rm ext}$  respectively, where $P_{\rm ext}$ is the external pressure, and all quantities have 
the usual meaning, consistent with those used by
\citet{springel05}. The pair-wise nature of the force summation over
the SPH neighbors ensures that $P_{\rm ext}$ cancels for particles that are surrounded
by other particles. At the edge, where the term does not disappear, it mimics
the pressure contribution from a surrounding medium. The values of $P_{\rm ext}$ used in this study are $3 \times 10^{7} k_{\rm B} \: {\rm K \: cm^{-3}}$ for the Pop III.1 clouds, and $7.5 \times 10^{6} k_{\rm B} \: {\rm K \: cm^{-3}}$ for the Pop III.2 clouds, where $k_{\rm B}$ is the Boltzmann constant. These confining pressures are similar to the internal pressure of the clouds in each case (see the temperatures and densities in \S \ref{sec:ics} below).

Finally, we have added
a treatment of primordial gas chemistry and cooling to the code. To model the
thermal evolution of the gas, we use an operator-split formalism, which treats
the effects of radiative and chemical heating and cooling separately from 
compressional heating. The influence of radiative and chemical heating and 
cooling on the thermal energy of each particle can thus be formulated as an
ordinary differential equation (ODE), which can then be solved simultaneously
with the chemical rate equations (also ODEs) that describe the chemical 
evolution of the gas. As \citet{tao09} have previously discussed, the strong coupling 
between the chemical and thermal evolution of high density primordial gas
that results from the importance of three-body H$_{2}$ formation heating 
and H$_{2}$ collisional dissociation cooling renders it vital to treat cooling and
chemistry simultaneously; treatments that do not do so will produce numerically 
stable results only with great difficulty. Full details of our chemical model and cooling
function are given in the Appendix.

%
%
%
%
%
%
%

\begin{figure*}[t]
	\centerline{
    		\includegraphics[height=3.0in]{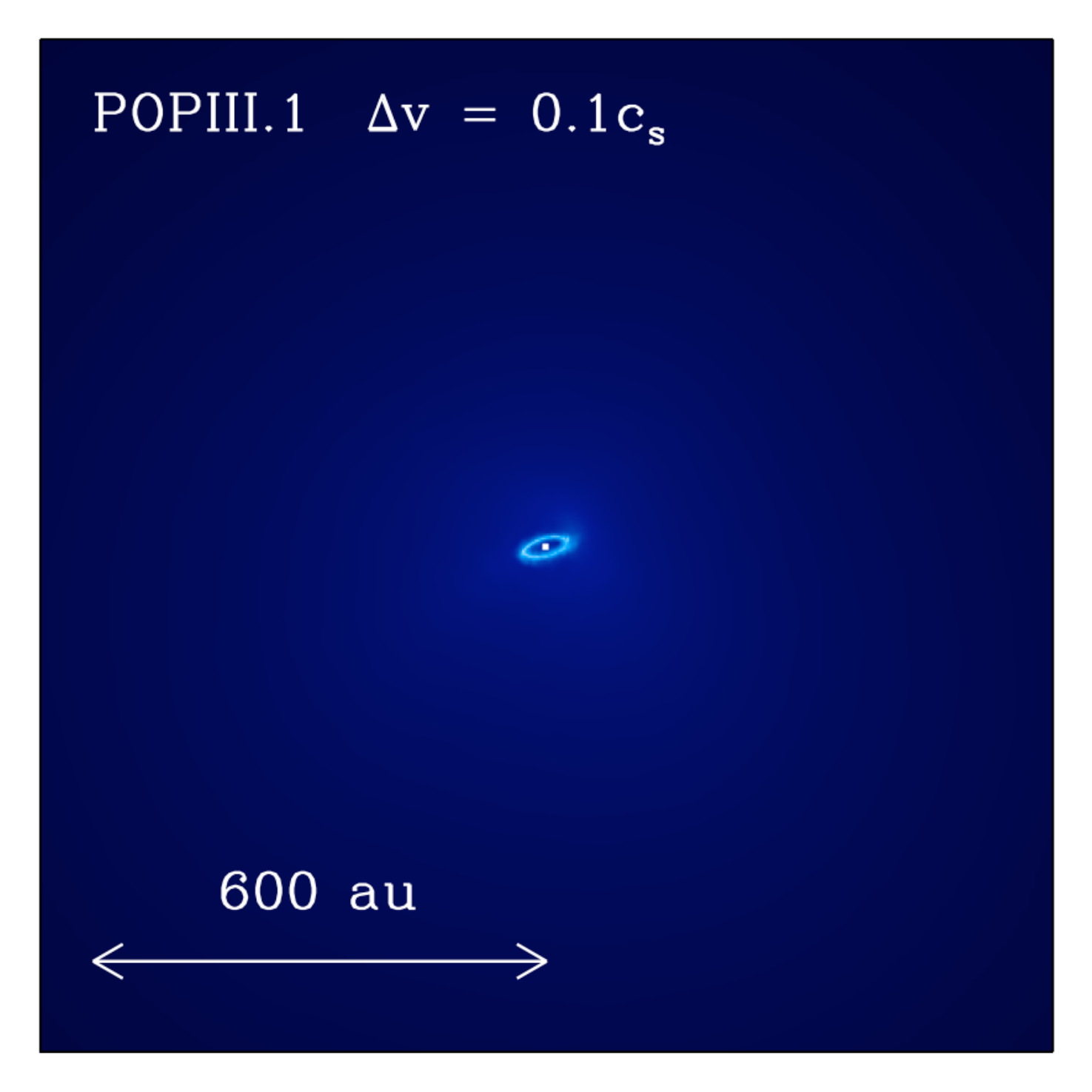}
		\includegraphics[height=3.0in]{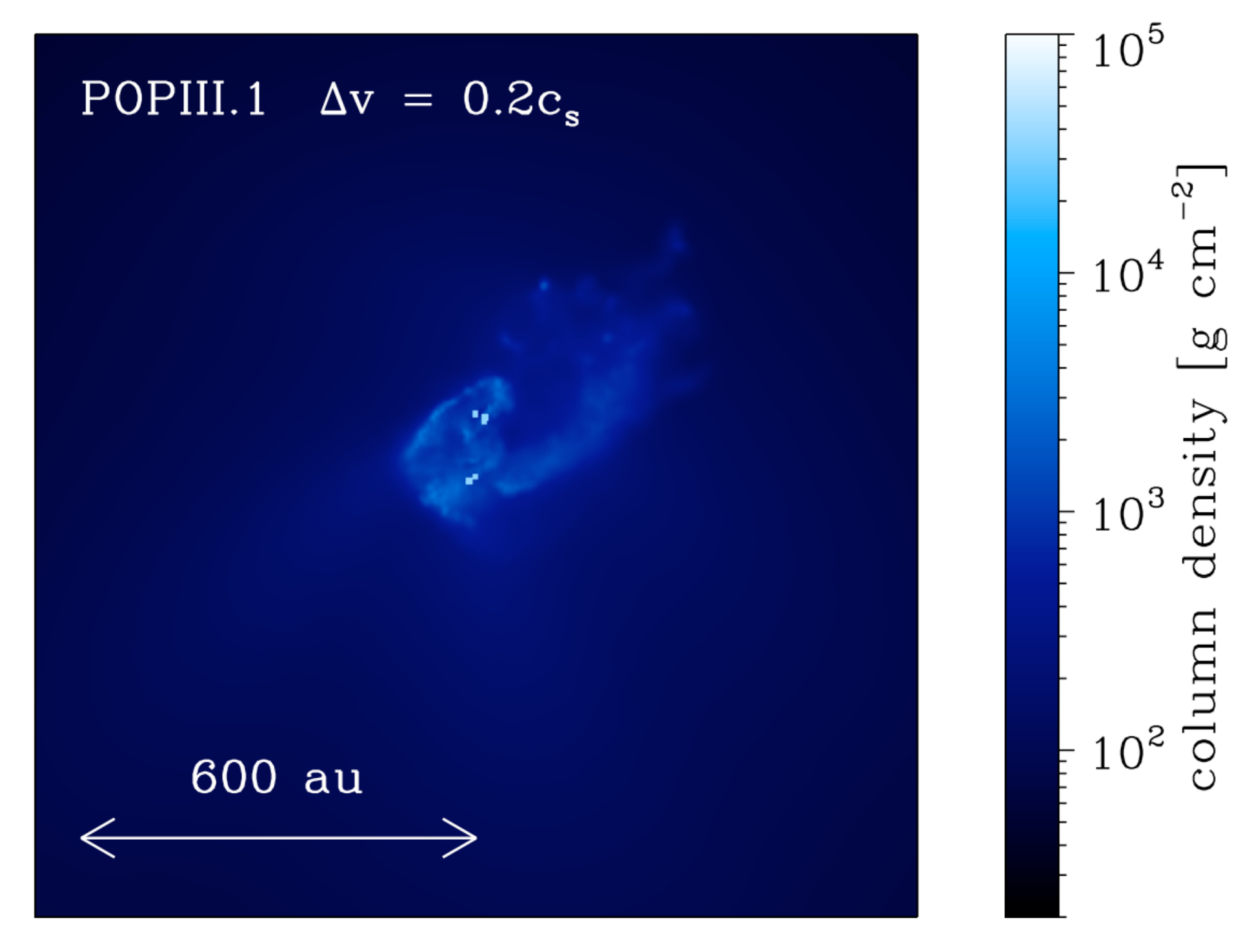}
	}
	\centerline{
		\includegraphics[height=3.0in]{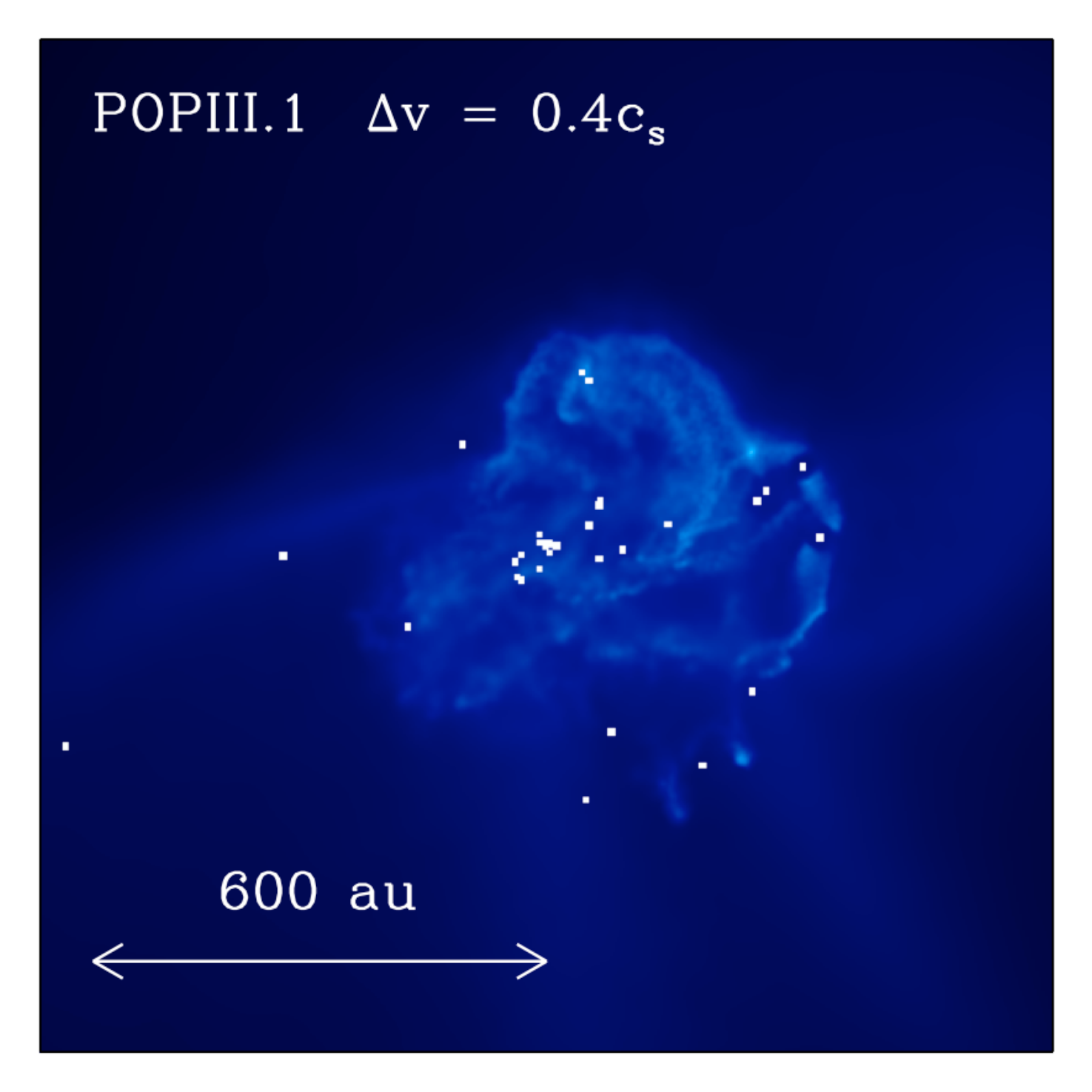}
		\includegraphics[height=3.0in]{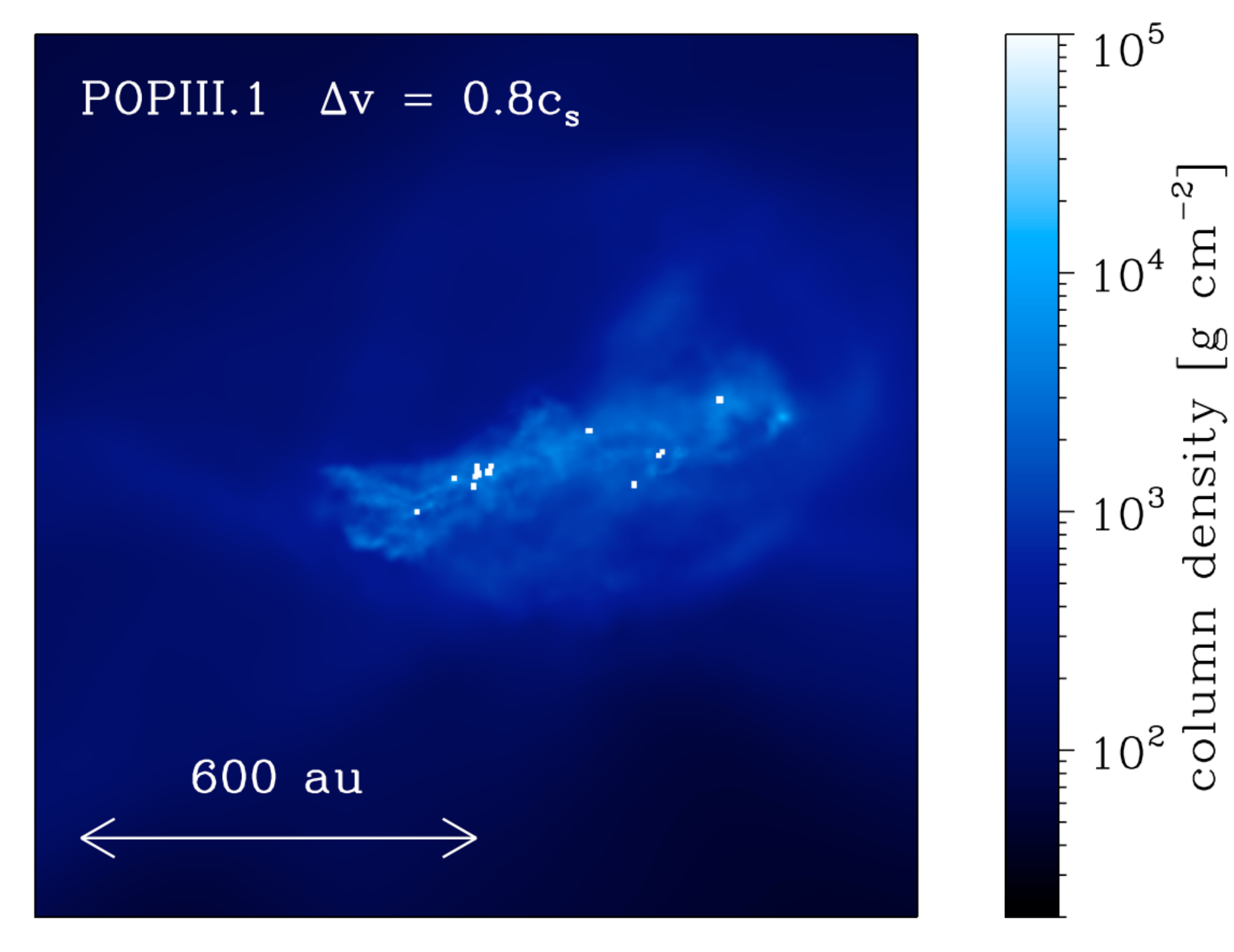}
	}
\caption{\label{fig:image3.1} Column density images showing the state of the Pop.\ III.1 clouds after they have converted 10 percent of their mass (100 \solmasp) into sinks. The sink particles are denoted in the images by the white dots, and in each case we centre the image on the first sink particle to form in the simulation.  For the run with the lowest level of turbulence ($\Delta v_{\rm turb} = 0.1 c_{\rm s}$) the initial turbulent 
velocity field provides the gas with enough angular momentum to produce a small disk around the
sink particle, but does not cause fragmentation of the gas. However, once the strength of the initial turbulent velocities is increased to as little as 0.2 times the initial sound speed in the gas, the turbulence induces fragmentation. At the point at which the simulations are shown here, the 0.1, 0.2, 0.4 and 
0.8~\cs runs have produced 1, 5, 31 and 15 sink particles, respectively. Note that the sink particles in this study have an accretion radius of 20 au.}
\end{figure*}

\begin{figure}[t]
	\centerline{		
    		\includegraphics[height=2.5in]{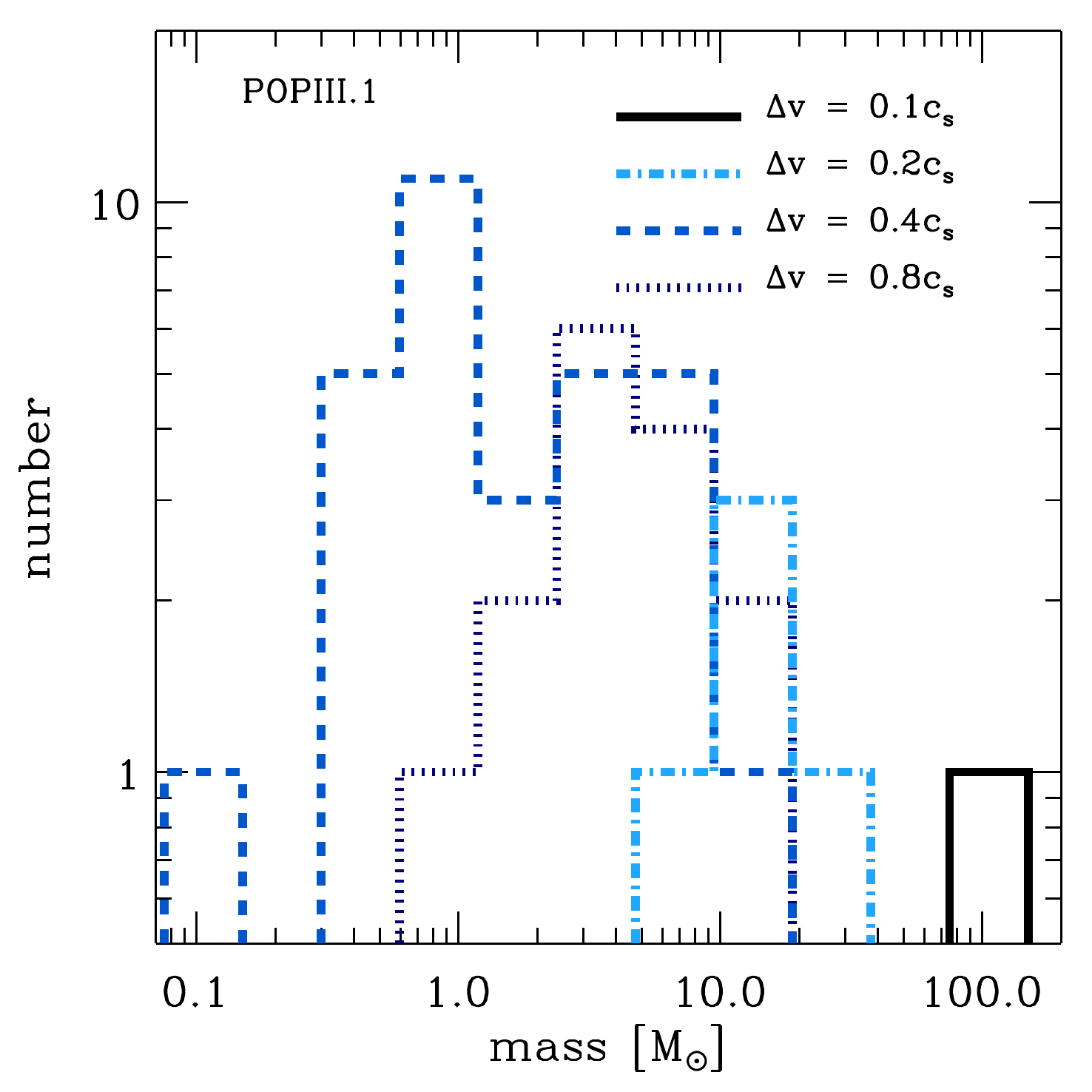}
		}
	\centerline{		
		\includegraphics[height=2.5in]{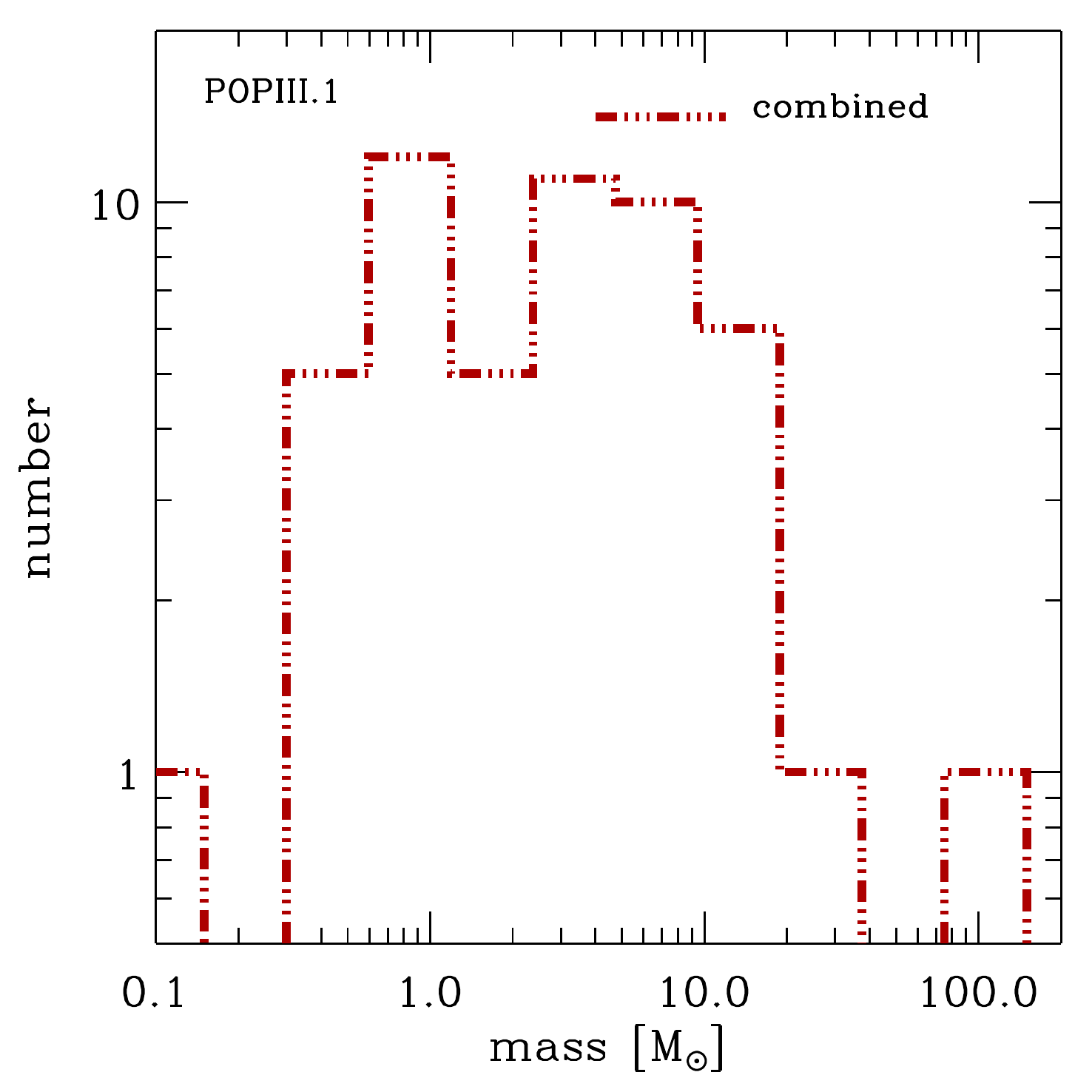}
		}
\caption{\label{fig:sinkmf} The top panel shows the mass functions from those Pop.\ III.1 simulations in which fragmentation occurs. In all cases the mass function is plotted at the point at which the total
mass of gas converted to sink particles is 100\solmasp. 
Note that as accretion is ongoing, and the system is still
young ($t \sim 1000$~yr), these will often not be the final masses of the sinks. The mass functions
in the individual simulations differ substantially, although the combined mass function, shown in the bottom panel, exhibits a broad and flat distribution between masses of 0.4 and 20 \solmasp.}
\end{figure}

\begin{figure*}[t]
	\centerline{
    		\includegraphics[height=3.4in]{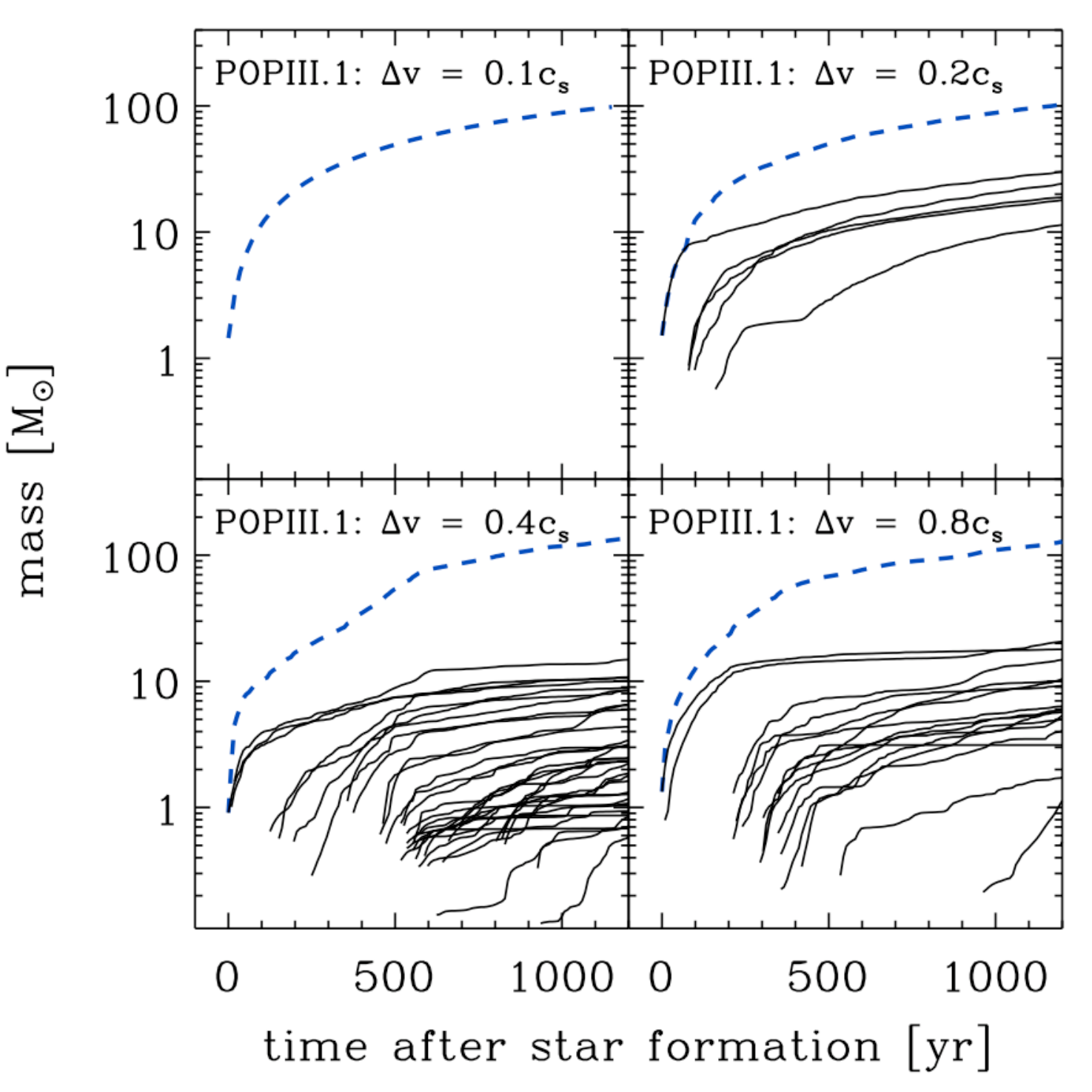}
		\includegraphics[height=3.4in]{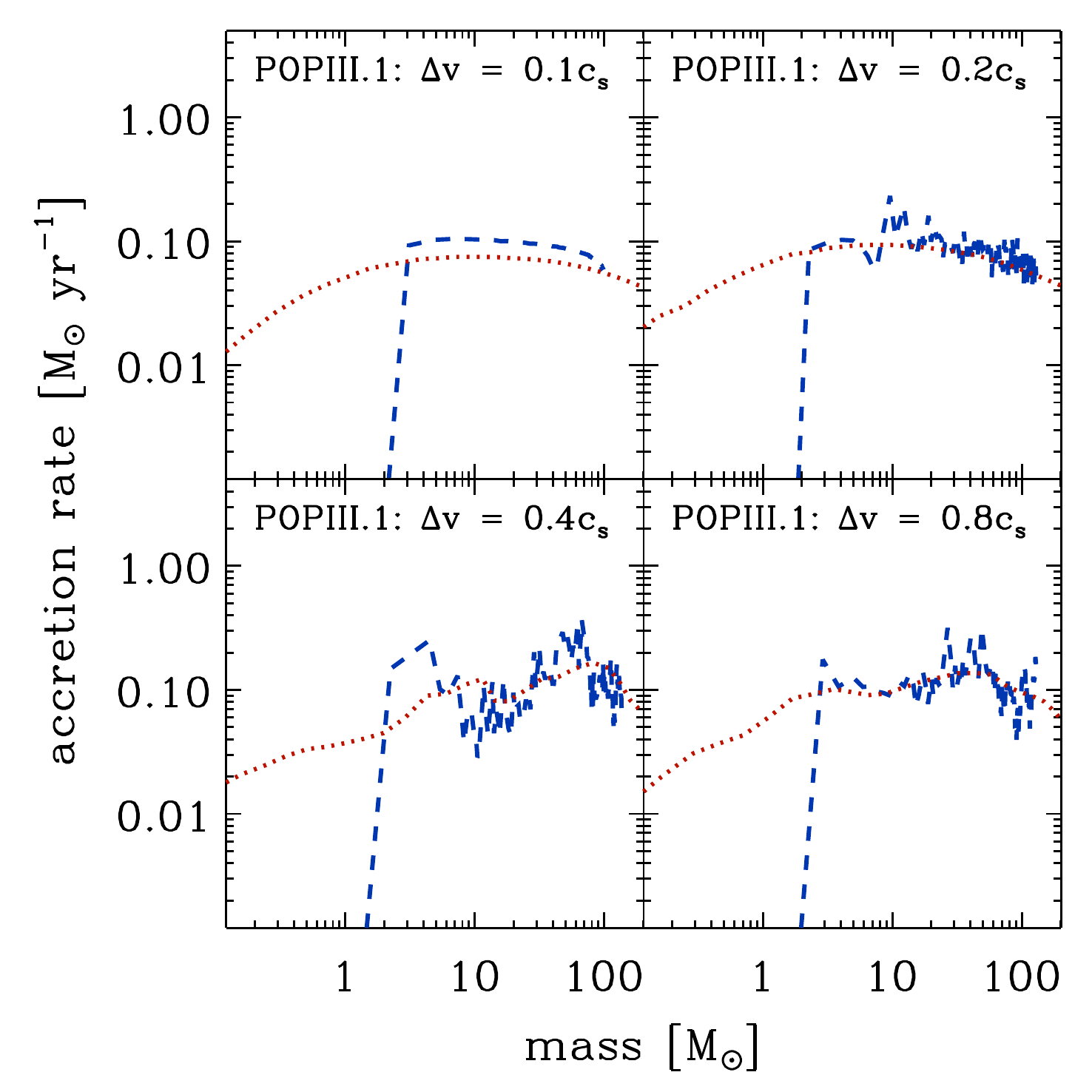}
 	}
\caption{\label{fig:maccp1} In the left-hand panel we show the mass evolution of the sink particles that form during the first 1200 years in the Pop.\  III.1 simulations, during which just over 10 percent of the initial gas mass is accreted. The solid (black) lines chart the mass of individual sink particles while the dashed (blue) lines show the mass evolution of the entire cluster of sink particles. The right-hand panel shows the associated mass accretion rate of the cluster, both as measured by summing over all sink
particles (blue dashed) and from an estimate based on the radial mass infall profile just before the formation of the first sink particle (red dashed).}
\end{figure*}

\begin{figure*}[t]
	\centerline{
		\includegraphics[height=3.in]{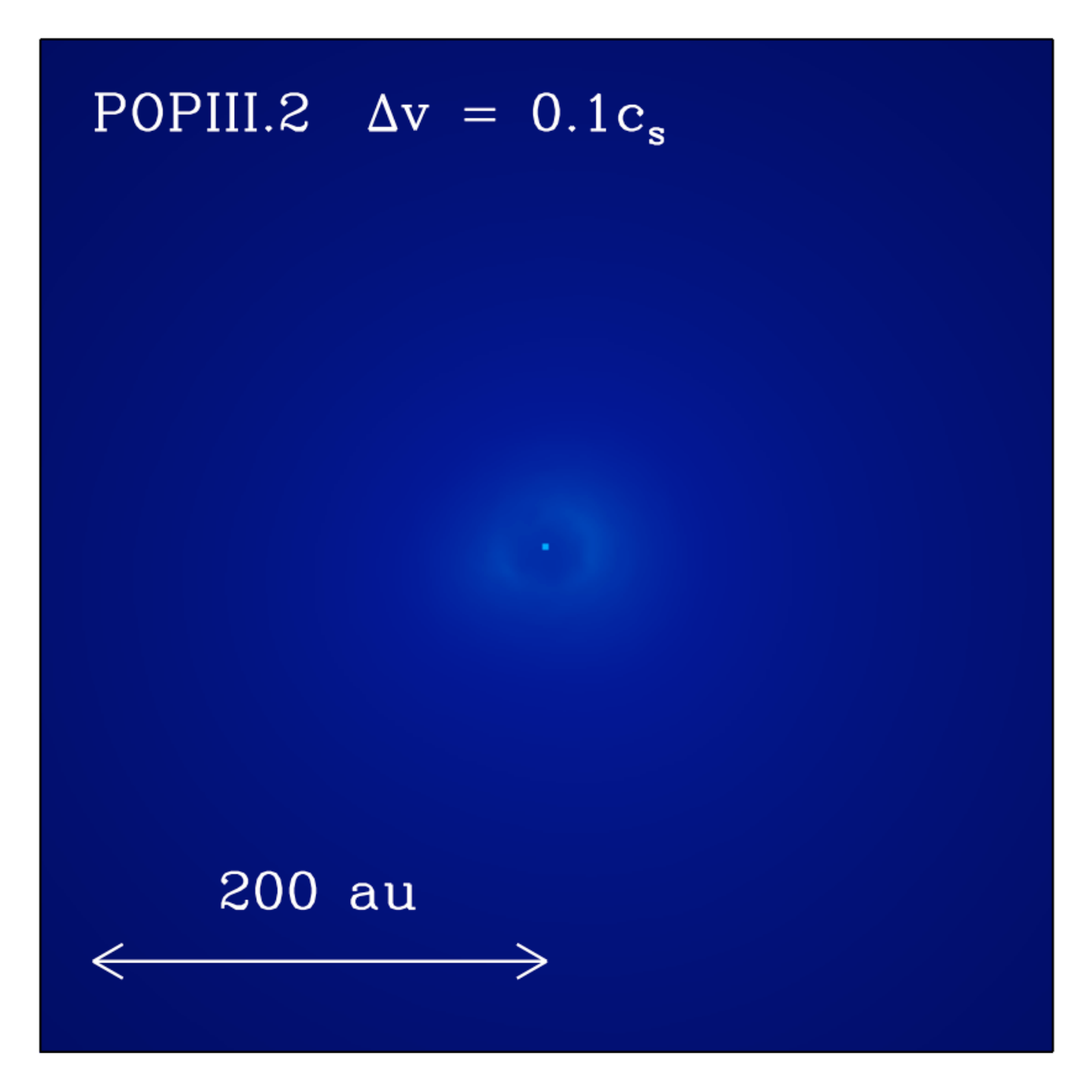}
		\includegraphics[height=3.in]{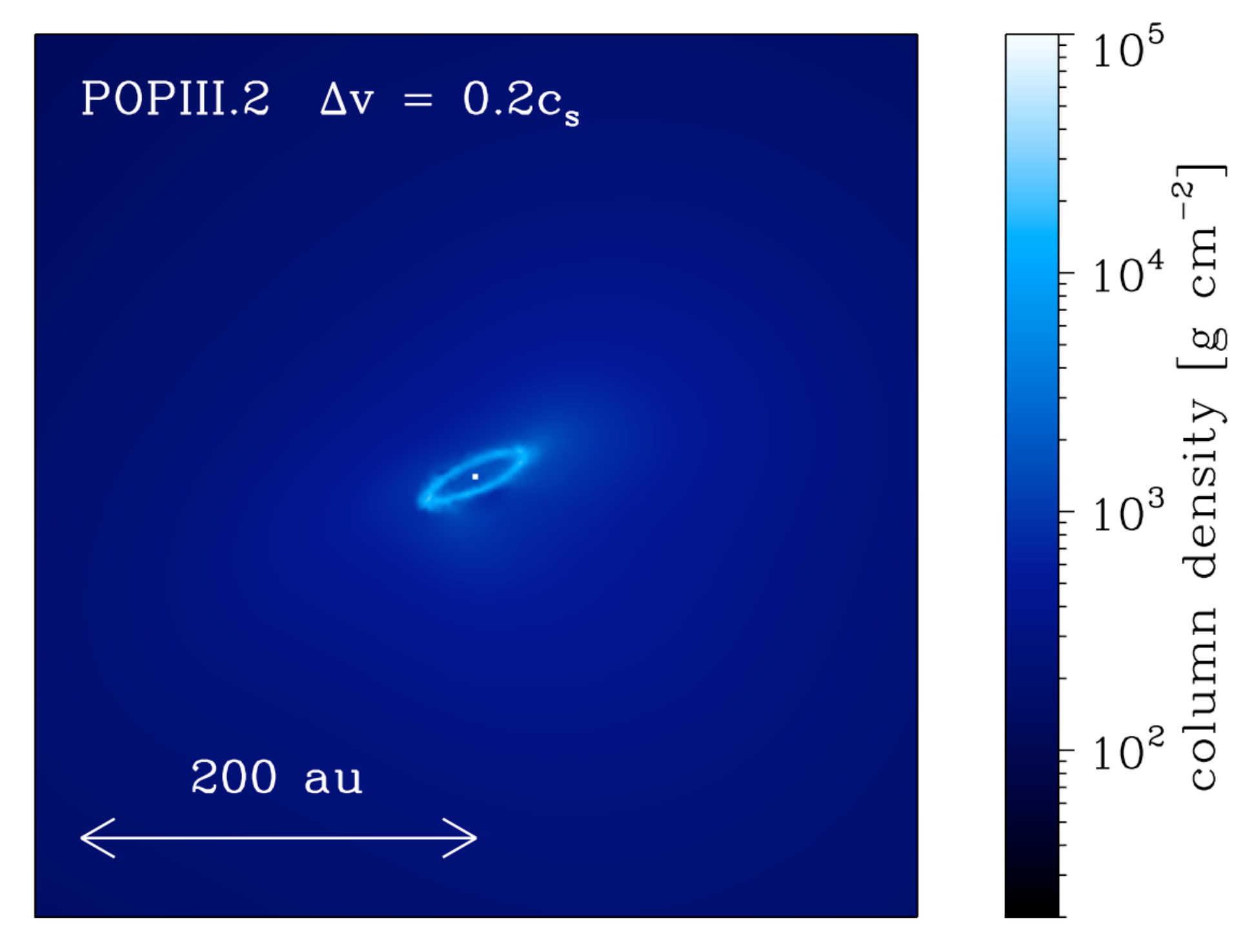}
	}
	\centerline{		
		\includegraphics[height=3.in]{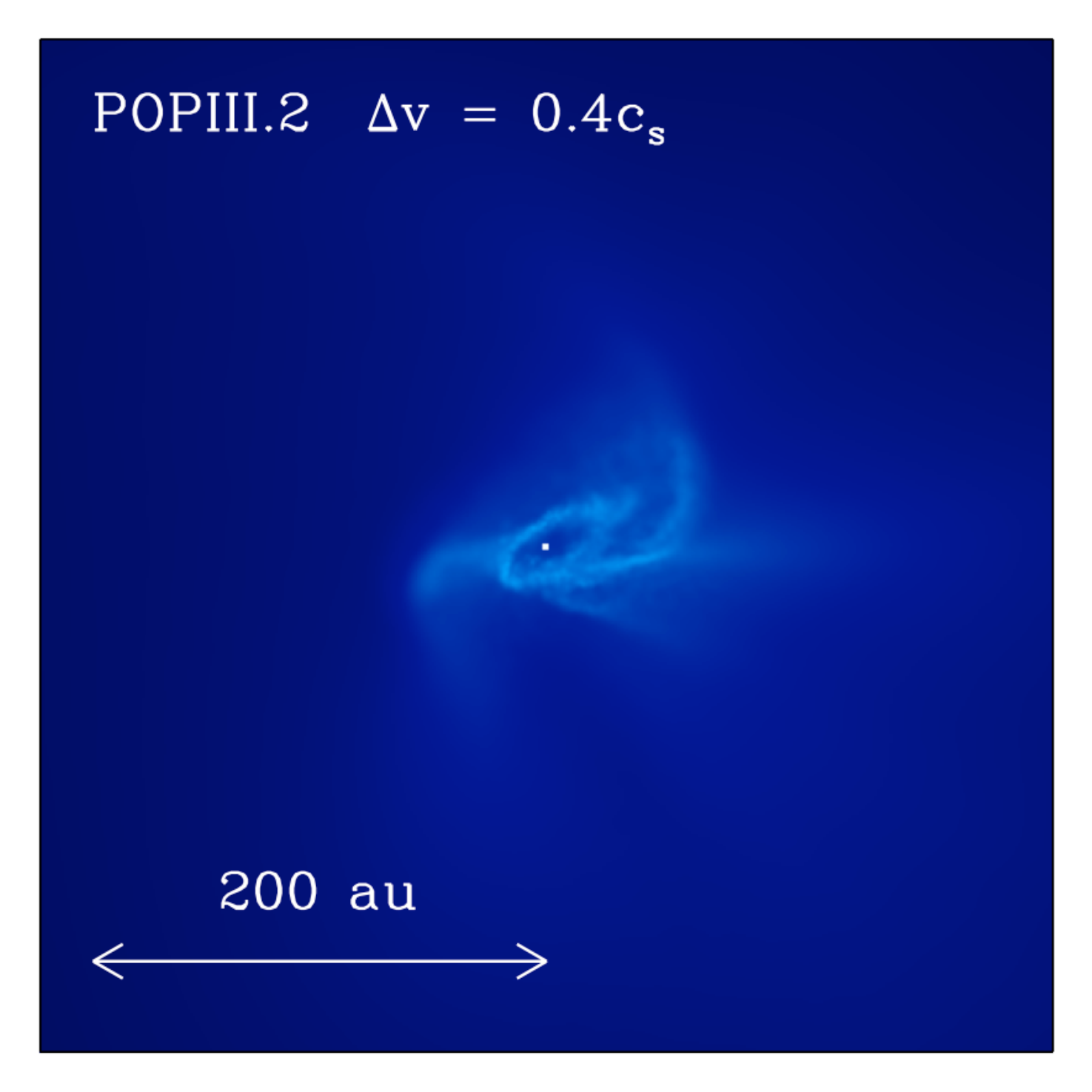}
		\includegraphics[height=3.in]{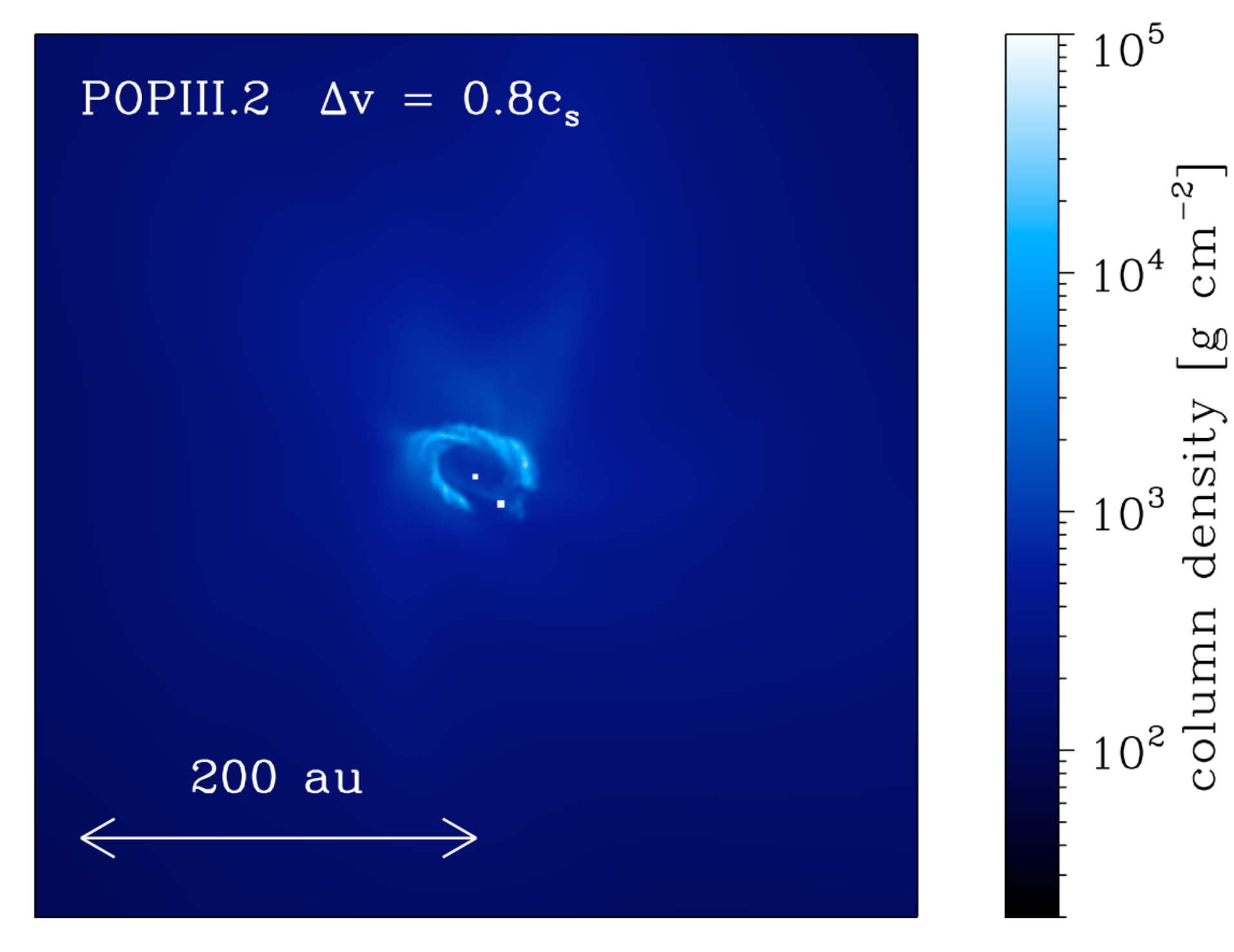}
	}
\caption{\label{fig:image3.2} Column density images showing the state of the $150 \: {\rm M_{\odot}}$
Pop.\ III.2 clouds after they have converted 10 percent of their mass (15 \solmasp) into sinks. Note that the scale in this figure differs from that in Fig~\ref{fig:image3.1}. The clouds exhibit a different behaviour from their Pop.\ III.1 counterparts. Although the clouds all form disks around their sink particles, due to the angular momentum contained in the initial turbulent motions, only the $\Delta v_{\rm turb} = 0.8 c_{\rm s}$ turbulent cloud has undergone fragmentation by this point in the simulation. Note that the sink particles in this study have an accretion radius of 20 au.}
\end{figure*}

\begin{figure}[t]
	\centerline{\includegraphics[height=3.0in]{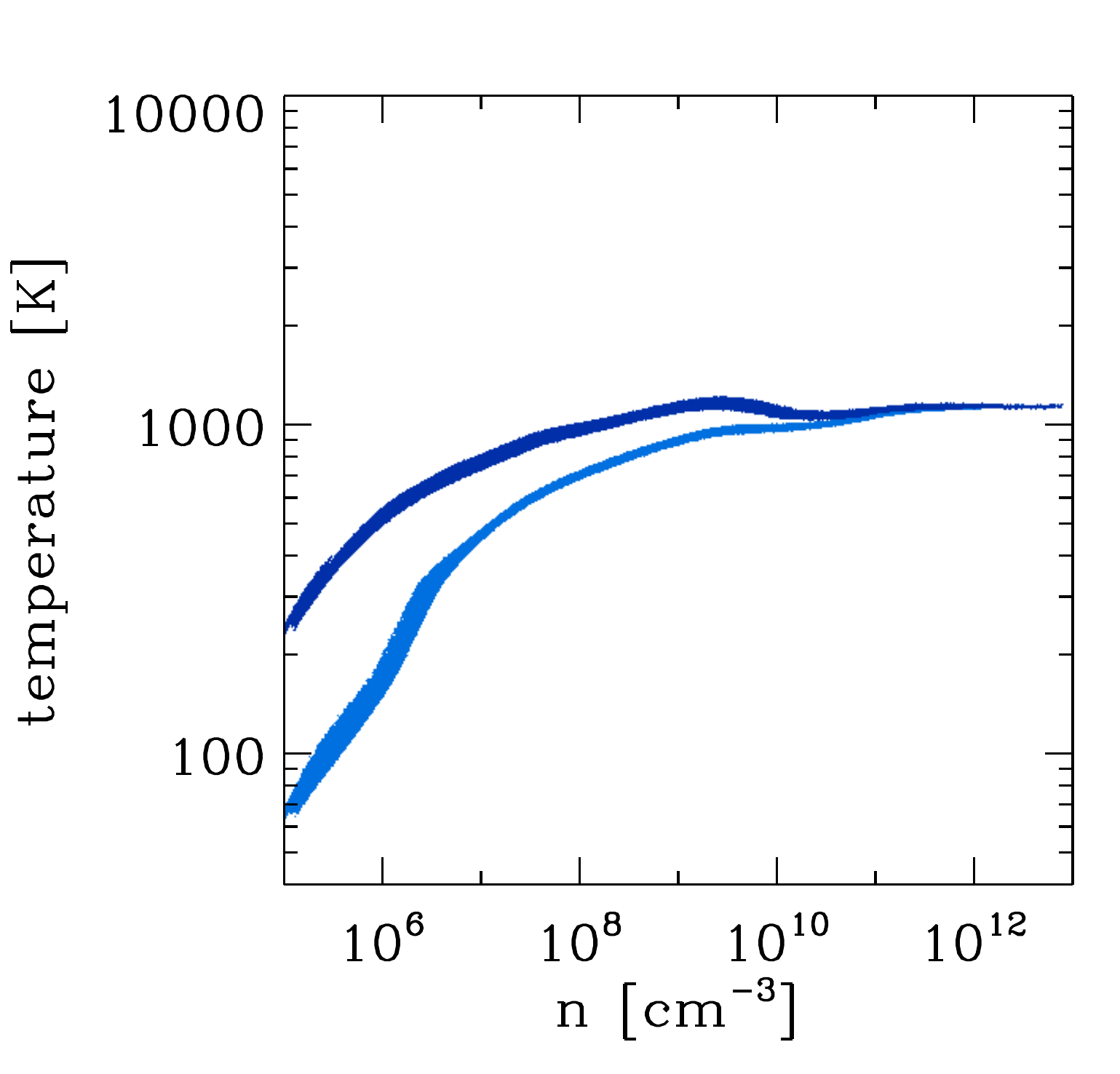}}
\caption{\label{fig:rhotevol} Temperature as a function of number density for the Pop.\ III.1 (dark blue) and Pop.\ III.2 (light blue) $\Delta v_{\rm turb} = 0.1\,c_{\rm s}$ simulations. In both cases, the curves denote the state of the cloud at the point just before the formation of the sink particle. }
\end{figure}

\begin{figure*}[t]
	\centerline{
    		\includegraphics[height=3.in]{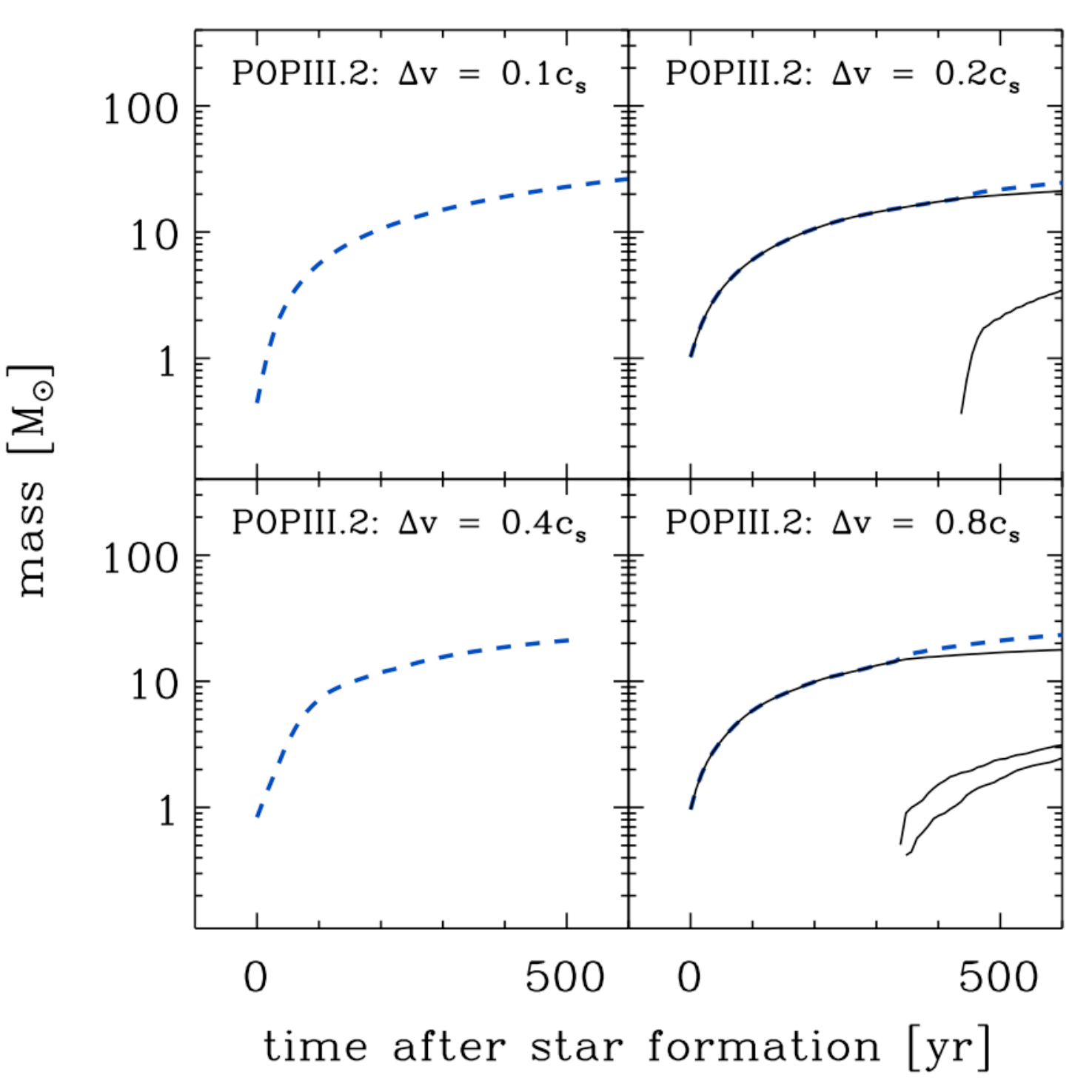}
		\includegraphics[height=3.in]{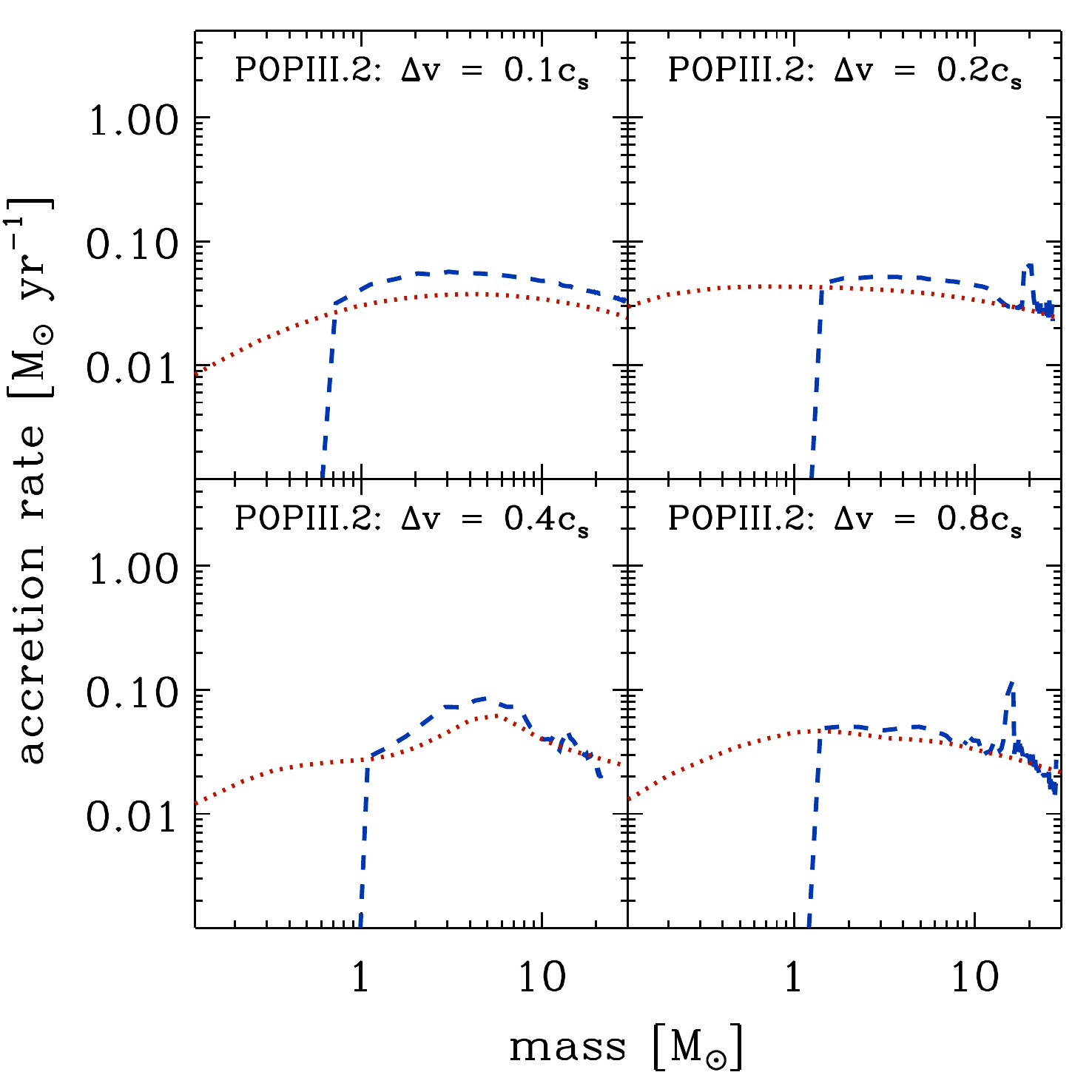}
 	}
\caption{\label{fig:maccp2} As Fig.~\ref{fig:maccp1}, but for the $150 \: {\rm M_{\odot}}$
Pop.\  III.2 simulations. Again, the evolution is plotted until slightly more than 10 percent of the cloud's mass has been accreted, which in these cases occurs after roughly 600 yr.}
\end{figure*}

\begin{figure}[t]
	\centerline{\includegraphics[height=3.0in]{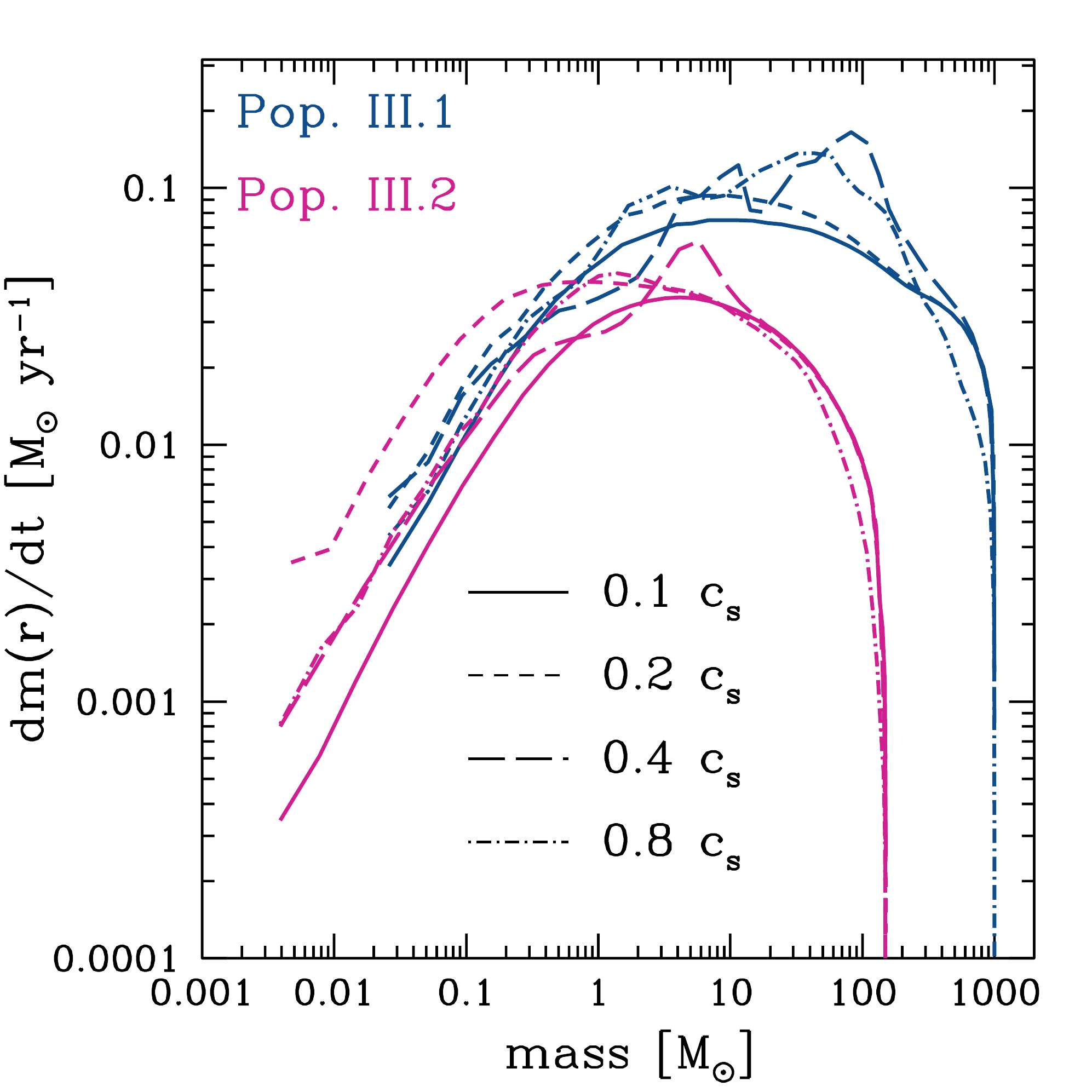}}
\caption{\label{fig:mdotcloud} Accretion rates as a function of enclosed gas mass in the Pop.\ III.1 
(upper lines; blue) and Pop.\ III.2 (lower lines; magenta) simulations, estimated as described in 
Section~\ref{pop31}. Note that the sharp decline in the accretion rates for enclosed masses close to the 
initial cloud mass is an artifact of our problem setup; we would not expect to see this in a realistic
Pop.\ III halo.}
\end{figure}

\begin{figure}[t]
    		\centerline{\includegraphics[height=3.in]{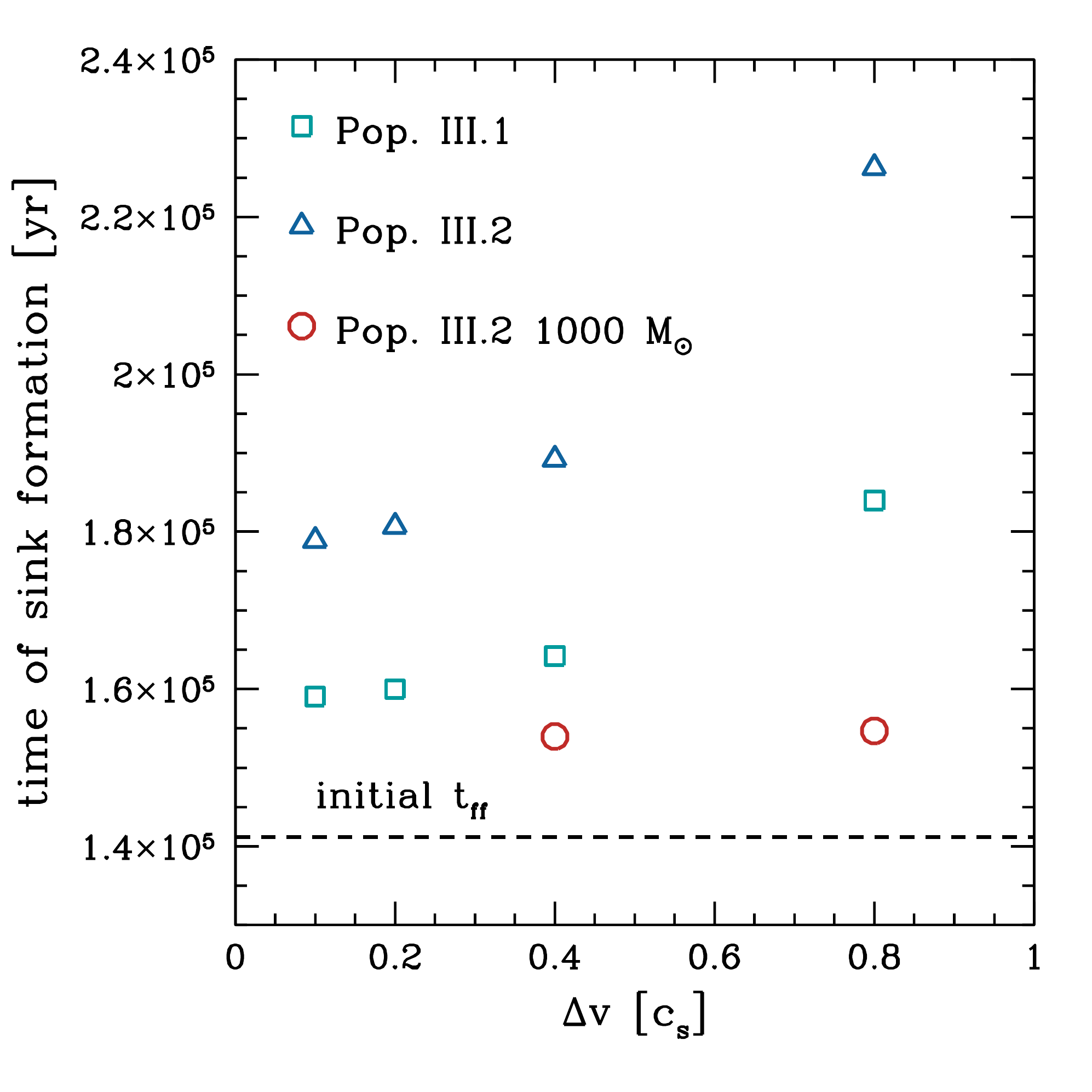}}
\caption{\label{fig:tsf} Time taken for the first sink particle to form in each simulation. For reference, the horizontal dashed line denotes the free-fall time of the clouds at their initial density.}
\end{figure}

\begin{figure*}[t]
	\centerline{
		\includegraphics[height=3.in]{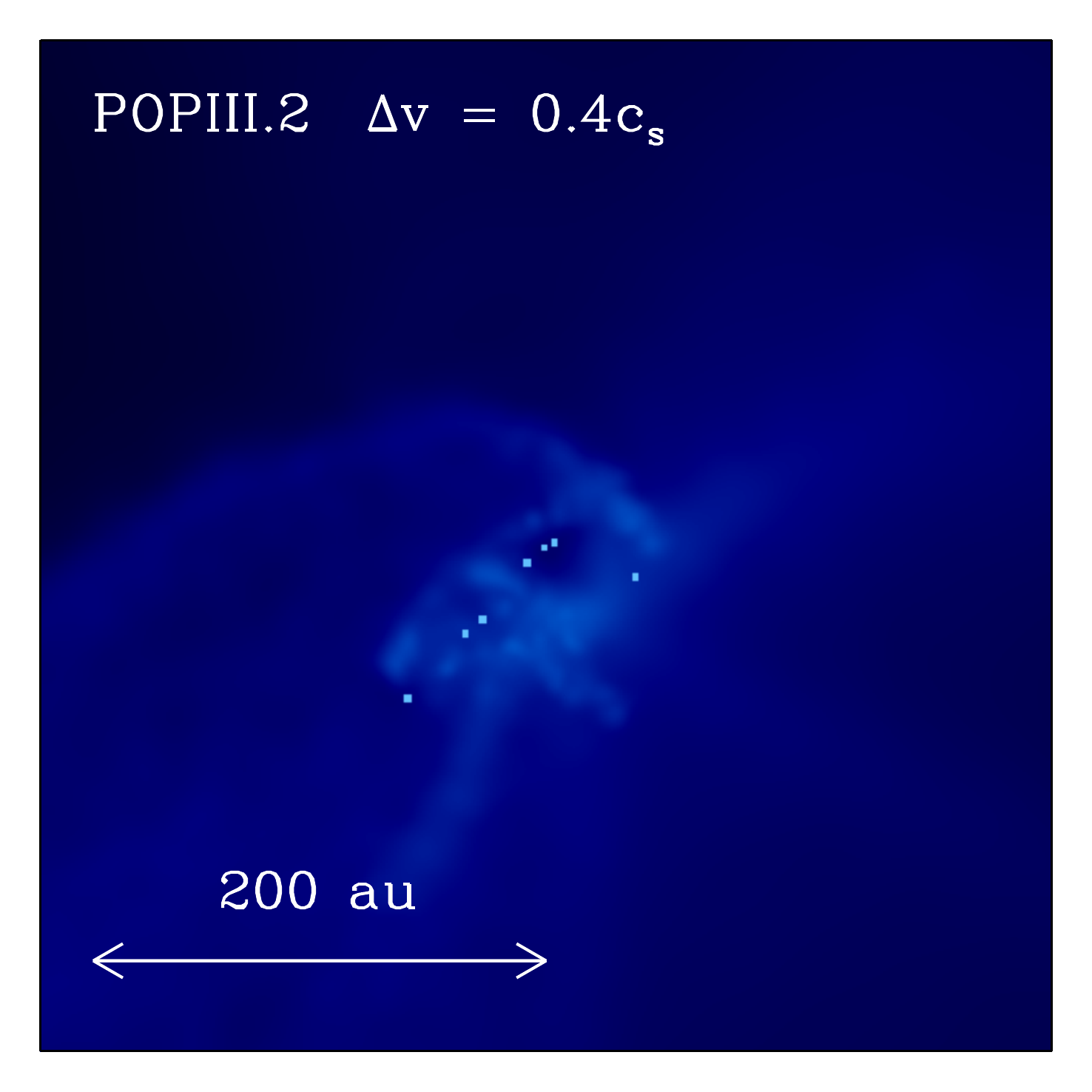}
		\includegraphics[height=3.in]{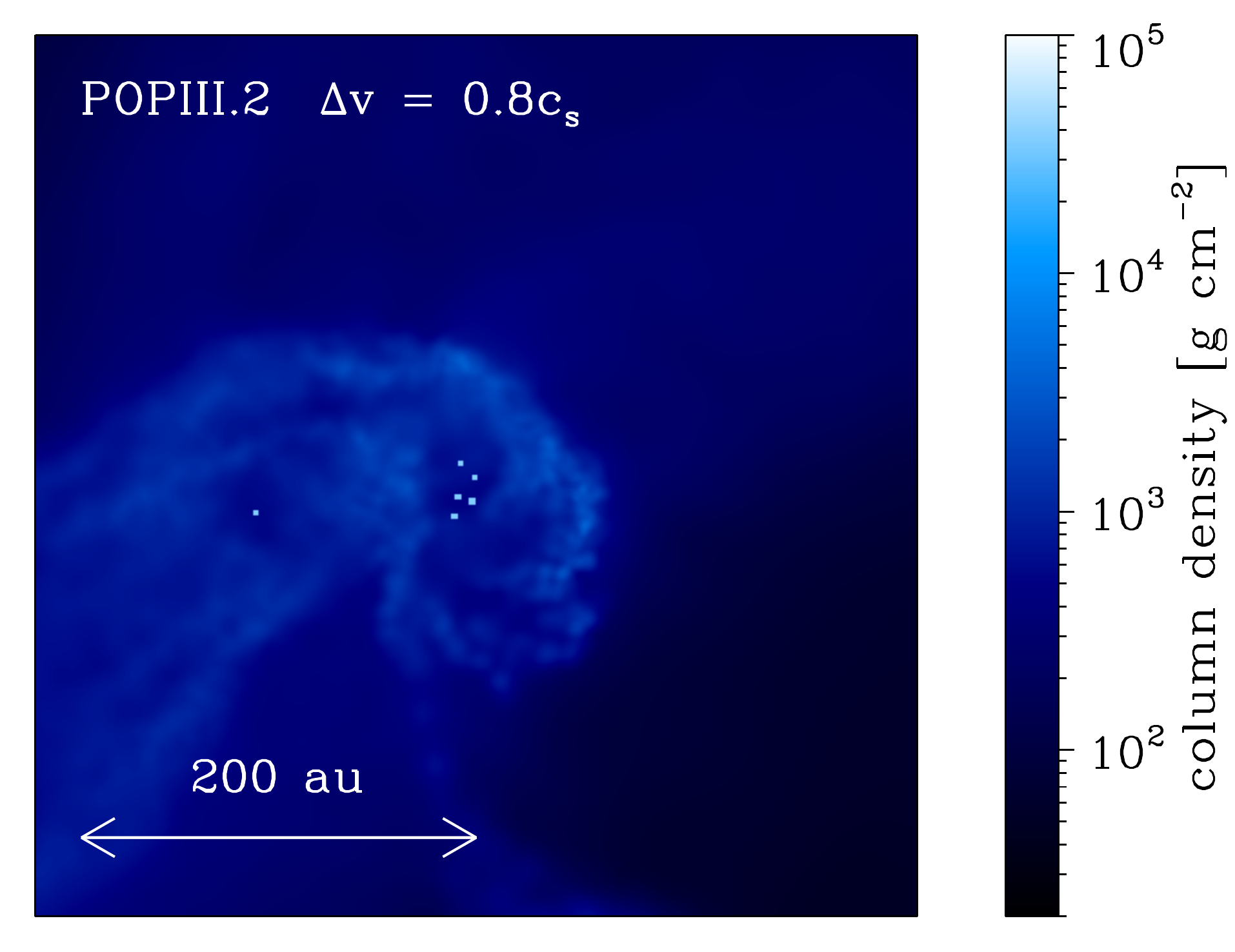}
	}
\caption{\label{fig:image3.2big} As Figs. \ref{fig:image3.1} and \ref{fig:image3.2}, but for the
1000~\solmas Pop.\ III.2 clouds after the sinks have accreted 10 percent of the total cloud mass (100 \solmasp). Note that the scale in this figure differs from that in Fig. \ref{fig:image3.1}.  Although these simulations have significantly more mass than those shown in Fig. \ref{fig:image3.2}, and are therefore initially more Jeans unstable, only a small amount of fragmentation is seen. The number of fragments
formed in each case is much smaller than in the corresponding Pop.\ III.1 runs. Note that the sink particles in this study have an accretion radius of 20 au.}
\end{figure*}

\begin{figure}[t]
	\centerline{		
    		\includegraphics[height=2.5in]{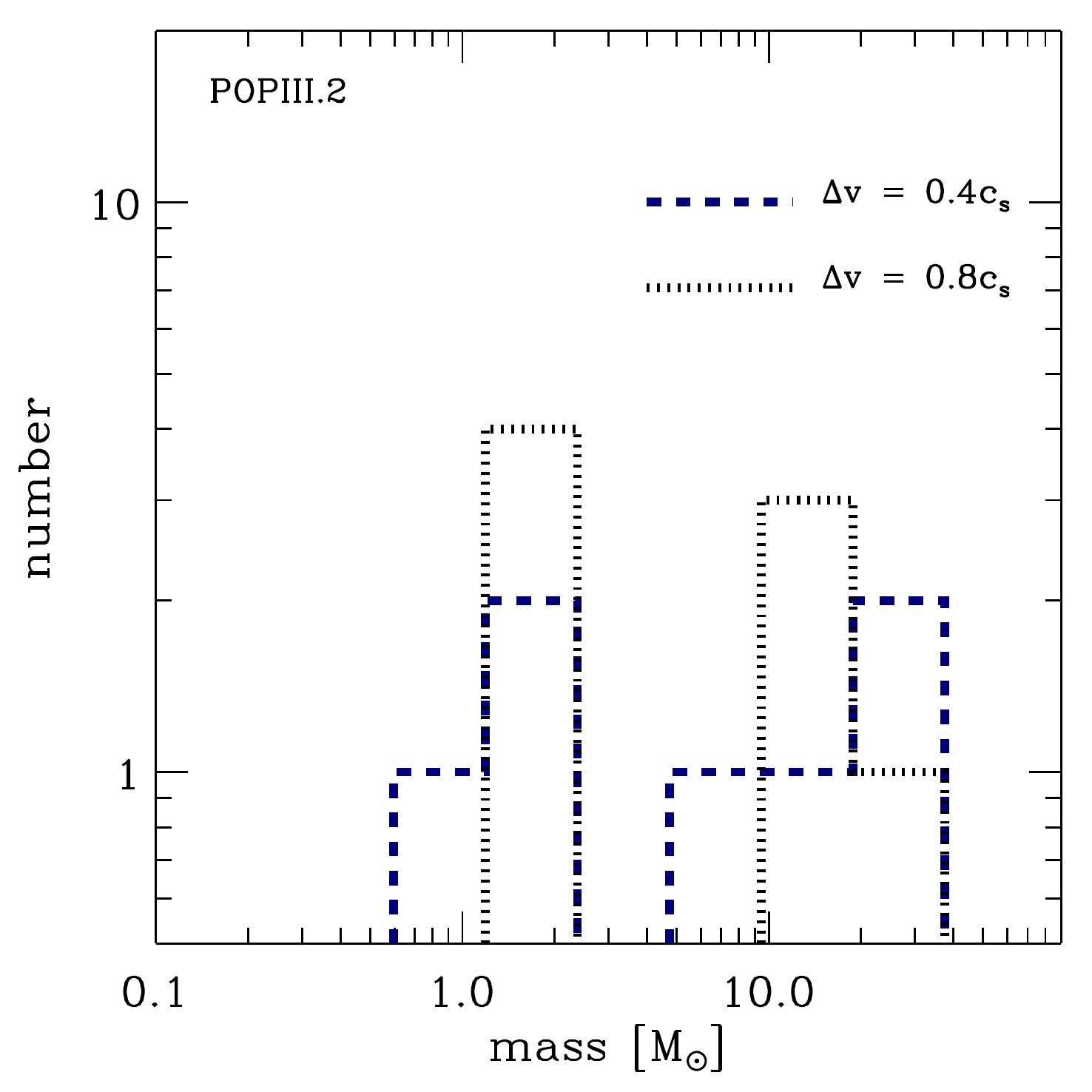}
		}
	\centerline{		
		\includegraphics[height=2.5in]{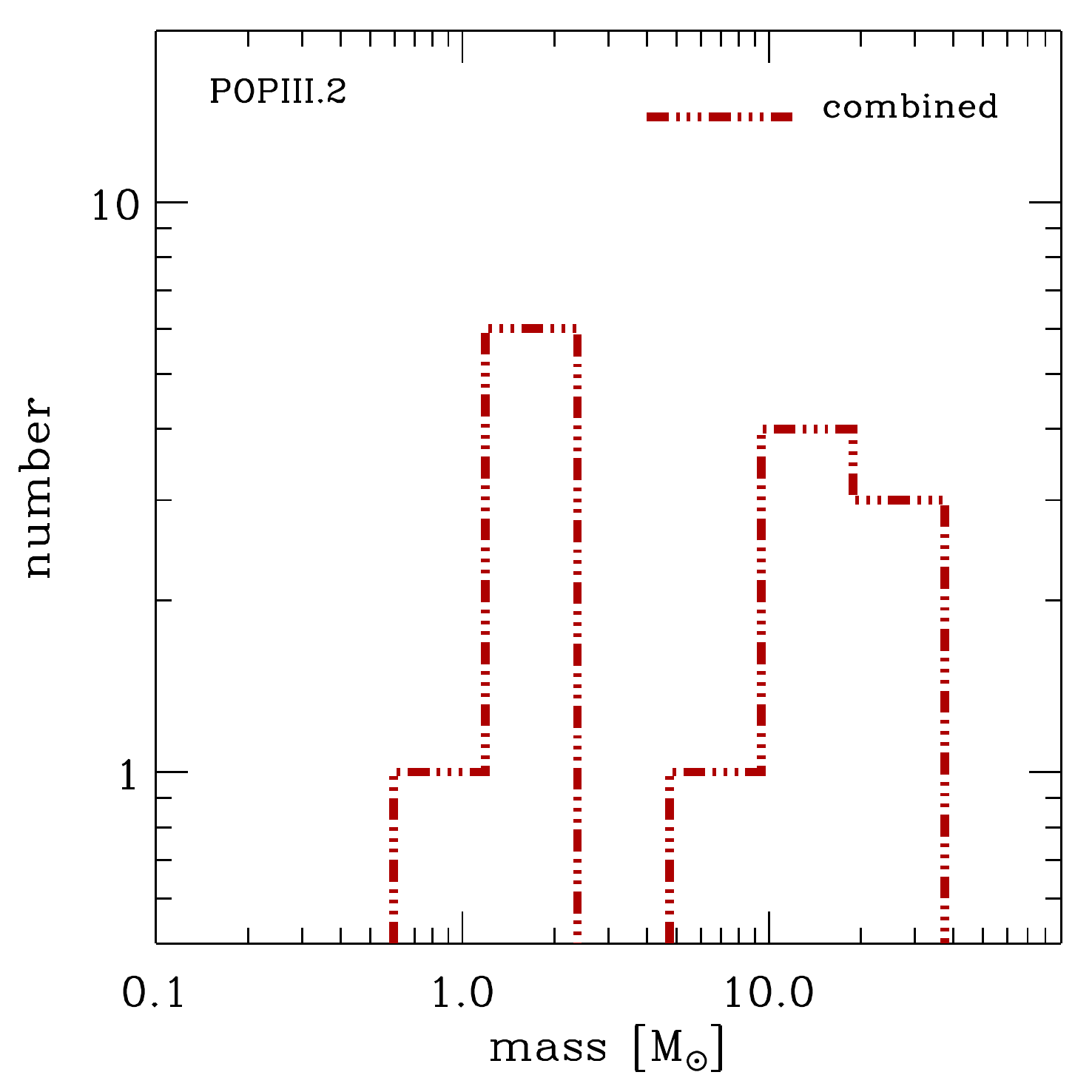}
		}
\caption{\label{fig:sinkmf32big} (top) Mass functions for our two 1000 \solmas Pop.\  III.2 clouds. As with the Pop.\  III.1 data in Fig. \ref{fig:sinkmf}, we show the mass function after 100 \solmas of gas has been accreted by the sink particles. Due to the reduced fragmentation in these calculations, the sink particles are on average more massive than their Pop.\  III.1 counterparts since they have less competition for the available mass. (bottom) The combined mass function for the two simulations.}
\end{figure}

\begin{figure*}[t]
	\centerline{
    		\includegraphics[width=2.5in]{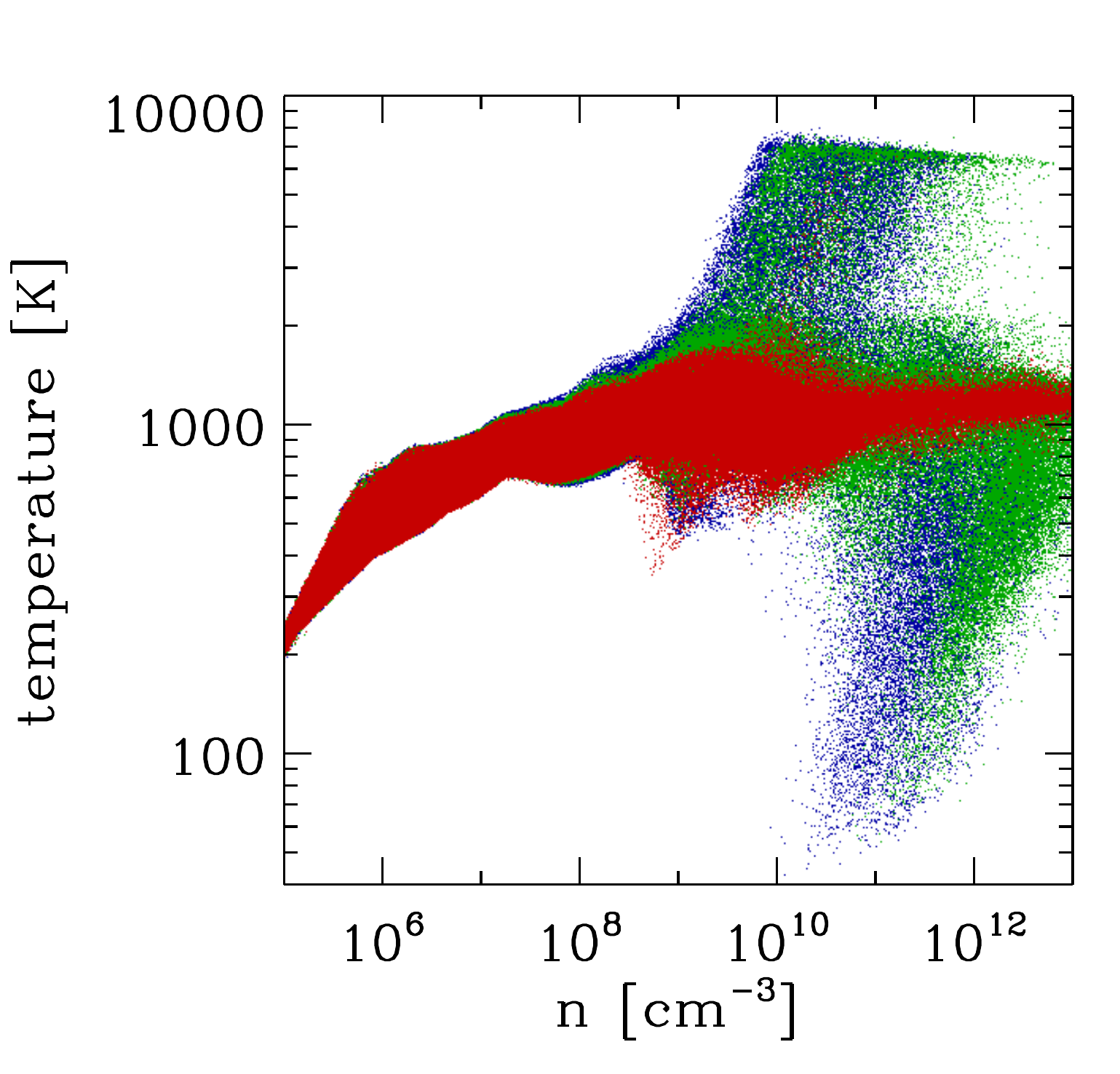}
		\includegraphics[width=2.5in]{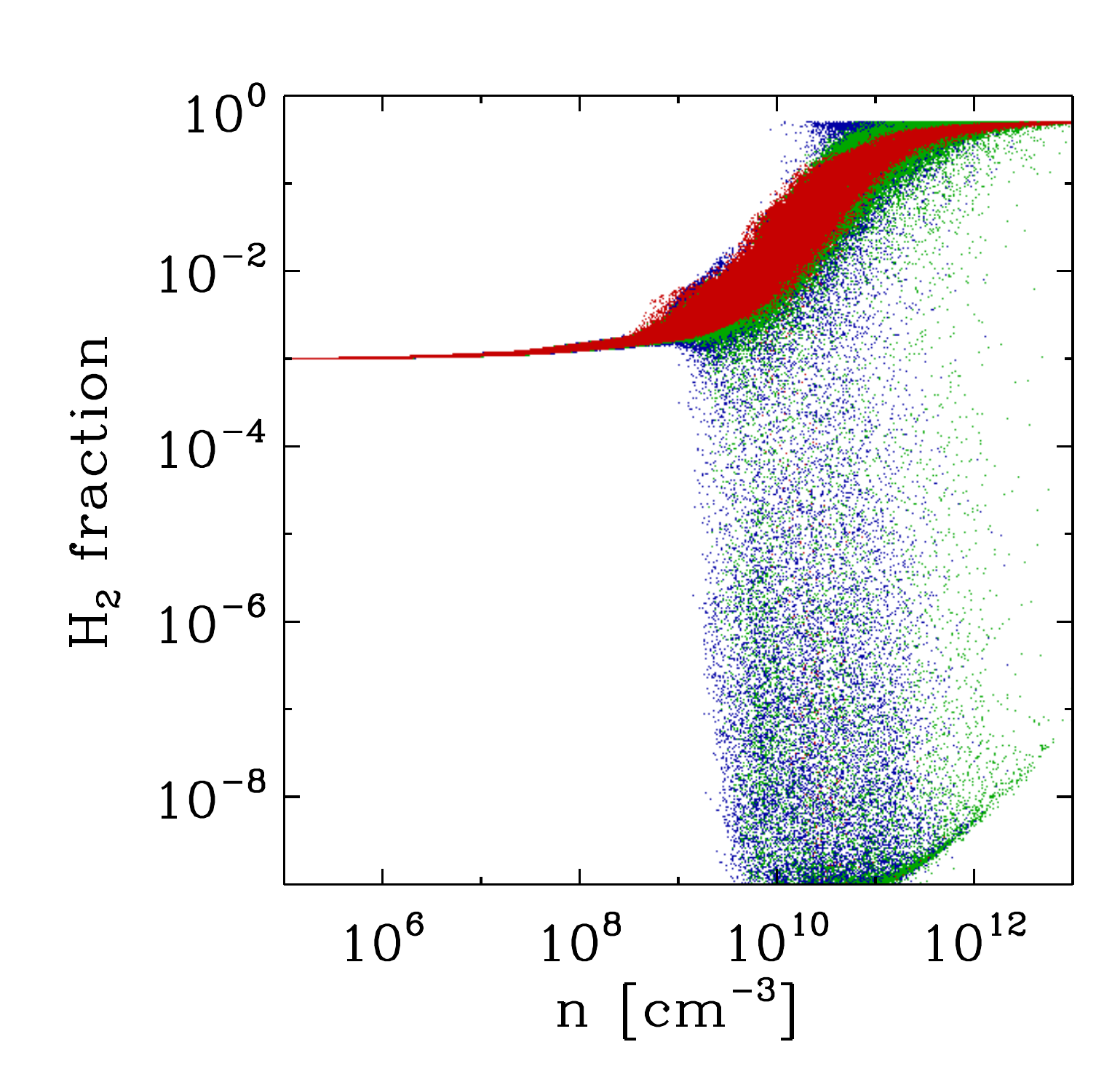}	
	}
	\centerline{
    		\includegraphics[width=2.5in]{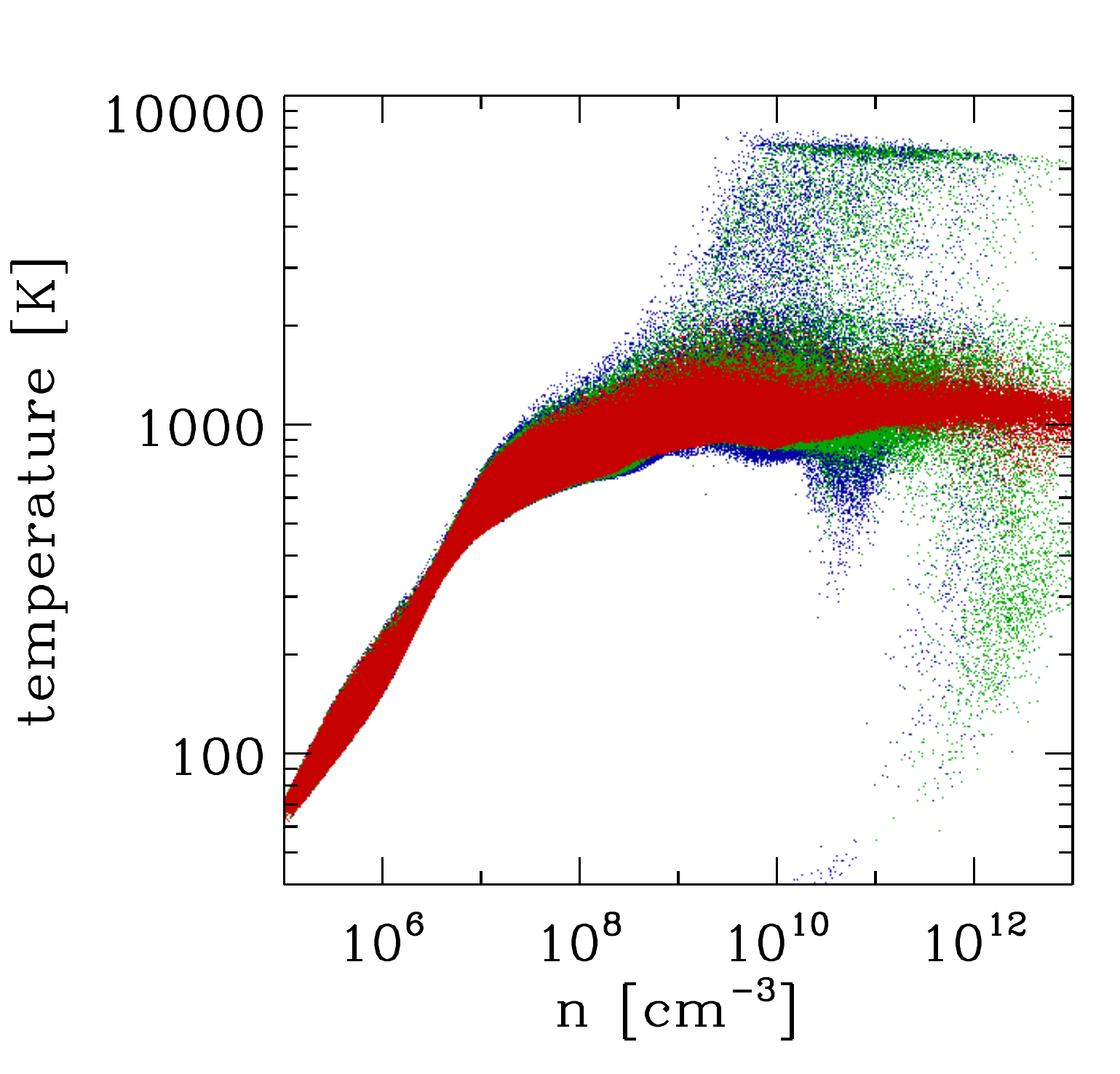}
		\includegraphics[width=2.5in]{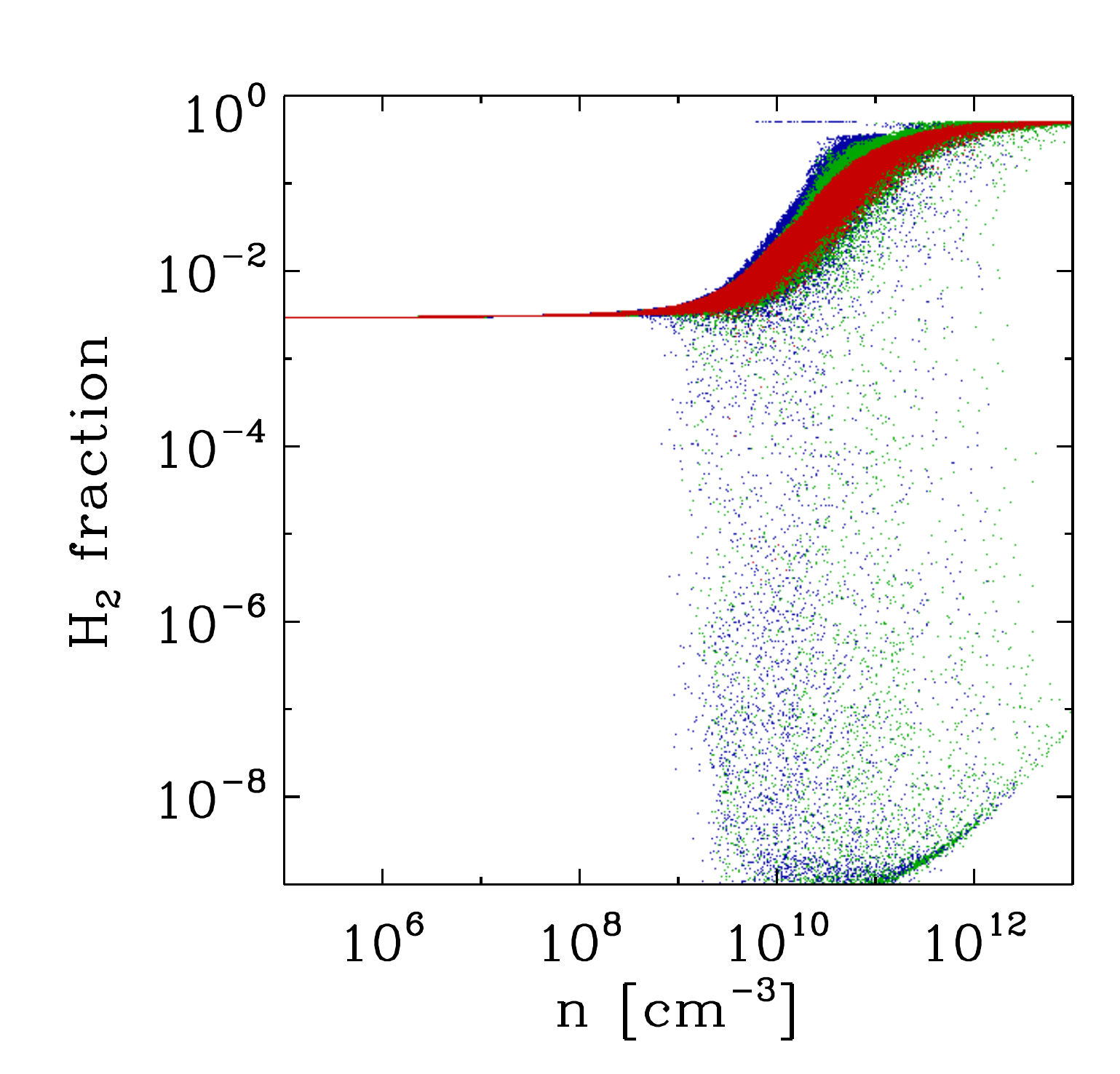}
	}
\caption{\label{fig:evol} Temperature and $\mHt$ fraction as a function of density in the Pop.\ III.1 (top) and 1000 \solmas Pop.\ III.2 (bottom) simulations in which $\Delta v_{\rm turb} = 0.4 c_{\rm s}$. Three stages in the evolution of the clouds are shown, corresponding to when the first sink forms (red), when there is 50 \solmas in sink particles (green) and finally when there is 100 \solmas in sink particles (blue).}
\end{figure*}

\begin{figure*}[t]
	\centerline{
    		\includegraphics[width=3.5in]{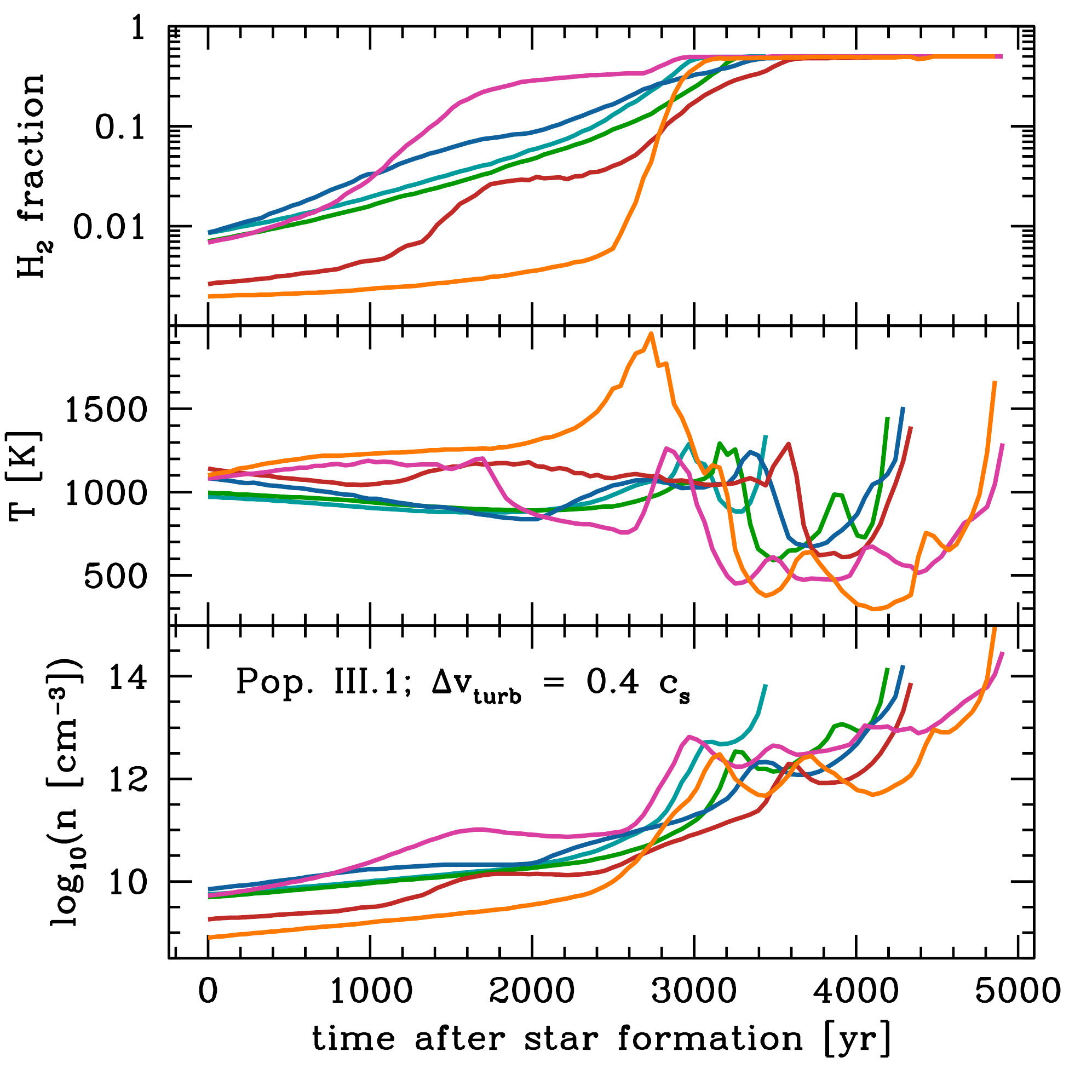}
		\includegraphics[width=3.5in]{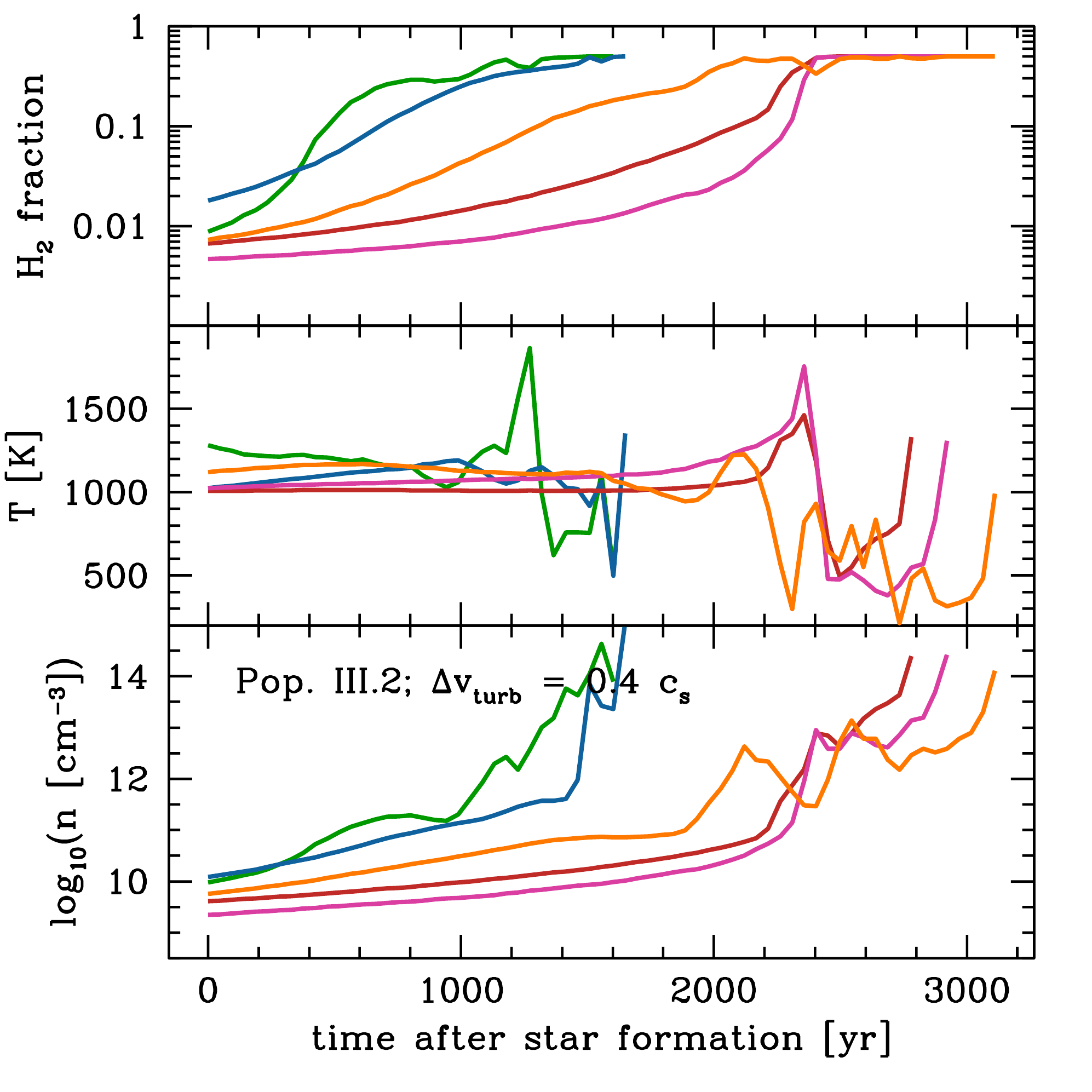}	
	}
\caption{\label{fig:ptrace} Particle trajectories showing the physical conditions in the gas leading up to the formation of sink particles in the 1000 \solmas Pop.\ III.1 and Pop.\ III.2 clouds, from the simulations with $\Delta v_{\rm turb} = 
0.4 \: {\rm c_s}$. The quantities are calculated by averaging the density, temperature and $\mHt$ over the 50 nearest neighbours of the particles which eventually become sink particles. Their evolution is shown from the onset of the star formation in the cloud (i.e.\ from the formation of the first sink particle), up to the point at which they themselves are turned into sink 
particles.}
\end{figure*}

\section{Initial conditions}
\label{sec:ics}

Since the aim of this study is to investigate the fragmentation properties of primordial gas, we use initial conditions that allow us to perform a controlled numerical experiment. The clouds start as unstable Bonnor-Ebert (BE) spheres into which we inject a subsonic turbulent velocity field. The BE sphere is made by allowing a gravitationally stable cloud of gas to evolve under its own self-gravity until the system has settled into a stable, centrally condensed configuration. During this initial settling phase, the gas temperature is held constant and the chemical evolution is not followed. On the scales of interest in this study, gravitational forces from the dark matter are negligible compared to the self-gravity of the
gas, and so for simplicity, we do not include dark matter in our models.

By re-scaling the mass and temperature of the cloud, we are then able to choose initial conditions that are gravitationally unstable, and which are similar to the initial conditions for the Pop.\ III.1 and Pop.\ III.2 star formation channels that occur in cosmological simulations. All of the simulations presented in this paper start with a maximum central cloud number density of $10^{5}$ cm$^{-3}$. The Pop.\  III.1 simulations start with an initial temperature of 300K and contain 1000 \solmas of gas. For the Pop.\ III.2 simulations, we adopt
an initial temperature of 75~K and examine two different initial masses. The first set of Pop.\ III.2
simulations contain 150 \solmas of gas, chosen such that they have the same ratio of thermal to gravitational energy as the Pop.\ III.1 simulations, roughly 0.15. As such, these Pop.\ III.2 simulations contain the same initial number of Jeans masses of gas as the Pop.\ III.1 simulations. For a uniform sphere, the number of Jeans masses is given by ($E_{\rm therm}/|E_{\rm grav}|)^{-3/2}$, so these clouds have roughly 3 Jeans masses in the initial configuration. All else being equal, if these clouds were to evolve isothermally from this point on, they would have the same propensity to fragment as the Pop.\ III.1 clouds, since this is directly related to the initial ratio of gravitational to thermal energy. Note that we have chosen a slightly subvirial configuration for our initial setup, to ensure that the clouds are still able to collapse when the turbulent motions are included. In addition to these simulations, we also performed a second set
of Pop.\ III.2 simulations that start with the same gas mass as in the Pop.\ III.1 case (1000 M$_{\odot}$). In this case, the  Pop.\  III.2 simulations are initially more Jeans unstable than their Pop.\  III.1 counterparts, and therefore might be expected to fragment significantly more. 

In our Pop.\ III.1 simulations, we set the initial fractional abundances of H$_{2}$, H$^{+}$, HD and 
D$^{+}$ to $x_{\rm H_{2}} = 10^{-3}$, $x_{\rm H^{+}} = 10^{-7}$, $x_{\rm HD} = 3 \times 10^{-7}$ 
and $x_{\rm D^{+}} = 2.6 \times 10^{-12}$, respectively. Our values for $x_{\rm H_{2}}$, $x_{\rm H^{+}}$
and $x_{\rm HD}$ are typical of the values found at these densities in cosmological simulations of
Pop.\ III.1 star formation \citep[see e.g.][]{2008MNRAS.387.1021G}, and account for the fact that
the HD/H$_{2}$ ratio is elevated over the cosmological D/H ratio of $2.6 \times 10^{-5}$ 
\citep{mol08} owing to the effects of chemical fractionation \citep{g08}. In the case of D$^{+}$, fractionation is unimportant at our starting temperature, and so we simply set ${\rm D^{+}} / {\rm H^{+}} 
= 2.6 \times 10^{-5}$. In our Pop.\ III.2 simulations, we adopt the same initial H$^{+}$ and D$^{+}$ abundances, but set $x_{\rm H_{2}} = 3 \times 10^{-3}$ and $x_{\rm HD} = 3 \times 10^{-6}$, following
\citet{2008MNRAS.387.1021G}. In both the Pop.\ III.1 and Pop.\ III.2 simulations, we assumed that
all of the helium remained neutral, and set the initial abundances of all of our other tracked species
to zero.

Within the BE spheres, we impose a turbulent velocity field that has a power spectrum of $P(k) \propto k^{-4}$. We assume that the turbulence considered here has its origin in gravitationally driven flows that arise as the gas and dark matter virialize in mini-halos (Wise \& Abel 2007; Greif et al. 2008; Klessen \& Hennebelle 2010). As the gas is compressible in nature the turbulent velocity field will have a power spectrum that is somewhat steeper than the standard Kolmogorov (1941) description for incompressible flows. However, this deviation is small and we note  that the the ability of a cloud to fragment is only weakly dependent on the power spectrum of the turbulence \citep{ddetal04}. The three-dimensional root-mean-squared velocity in the turbulent field -- which we will refer to as $\Delta v_{\rm turb}$ -- is then scaled to some fraction of the sound speed \cs in the initial conditions. For the simulations presented here we use four different rms velocities: 0.1, 0.2, 0.4 and 0.8~$c_{\rm s}$. For an isothermal sound speed and an adiabatic index of $\gamma = 5/3$, the  
corresponding ratios of the turbulent to thermal energy are given by $1/3(\Delta v_{\rm turb}/c_{\rm s})^{2}$, yielding 0.0033, 0.0133, 0.0533 and 0.2133 respectively, for our chosen values of $c_{\rm s}$. In order to focus on the effects of the turbulence, we do not include any ordered rotation of the initial gas cloud. Note, however, that this does not imply that the initial angular momentum of the cloud is zero, since the imposed turbulent velocity field gives the cloud a small amount of angular momentum. Note that we only consider subsonic turbulence in this study since our clouds have only a few Jeans masses, and supersonic turbulence would unbind them. To study the effects of supersonic turbulence, one would have to look at clouds that are initially more Jeans unstable than those we study here.

The clouds in our study are all modeled using 2,000,000 SPH particles. Although this means that the mass resolution is higher in the 150 \solmas Pop.\  III.2 clouds than in the other simulations, the Jeans mass at the point where sink particles form is well resolved in every case (see below). For this study, it is more important that the turbulence in all simulations is evolved with the same resolution, hence our choice of a constant particle number throughout.

Sink particles are created once the number density of the gas reaches $10^{13}$ cm$^{-3}$, at which point the gas has a temperature of around 1200K. The corresponding Jeans mass at this density and temperature is 0.08 \solmasp. Our mass resolution in the Pop.\  III.1 and the 1000 \solmas Pop.\  III.2 clouds is $2 N_{\rm neigh} m_{\rm part} = 0.05$ \solmasp, where $N_{\rm neigh}$ is the number of neighbors employed for force evaluations (in our case 50), and $m_{\rm part}$ is the mass of an SPH particle. The 150 \solmas Pop.\ III.2 simulations have a mass resolution of 0.008 \solmasp. Once the candidate particle has passed the criteria described in Bate et~al.~(1995),
it is replaced by a sink particle that can accrete gas particles that fall within its accretion radius $r_{\rm acc}$, which we fix at 20 AU. Note that this radius is significantly larger than the Jeans radius at the point that the gas reaches the density threshold for sink creation, which is around 6 AU. As such, all the fragmentation that we capture in this study is well resolved and not lying close to the limits of our resolution; if any `artificial' fragmentation were to occur, the sink particle would immediately swallow the offending region and replace it with a single accreting point. We also prevent sink particles from forming within $2\, r_{\rm acc}$ of one another. This prevents the formation of sinks out of gas that in reality would by accreted by a neighboring sink particle before it could go into direct collapse by itself. Lastly, gravitational interactions between sinks, and between the
sinks and the gas, are softened in the standard way in Gadget 2, using a fixed softening parameter of 5 AU for the sinks, and a variable softening parameter for the gas particles that is equivalent to their smoothing length.

We note that the sink algorithm employed here does not allow sink particles to merge, in contrast to the `sticky' sinks introduced in Bromm et al.~(2002). As such, our simulations may be biased towards low masses. Small, secondary sinks could be driven towards more massive ones via dynamical friction and coalesce with them. This effect, which acts to reduce the resulting number of fragments, has been seen in AMR simulations of massive star forming regions in the
present-day universe (Krumholz et~al.\ 2007, 2009). 
However, given the size of our sink particles, it is unclear whether sink-merging is really applicable: the protostellar radius is at most expected to be 0.5 au (and then only for a short time, e.g. \citealt{ho2009}), which is significantly smaller than both our adopted accretion radius and the gravitational softening. In fact, in the simulations from present-day star formation in which extremely small sink particles are used -- along with little, or no softening --- merging events are found to be fairly rare (e.g. \citealt{Bate2009}). They do however occur, and this should be borne in mind when interpreting the results of this study. Also, given the size of our sink particles, we are unable to resolve any of the disks that would invariably form around the young protostars and it has been suggested that such structures would be unstable to fragmentation on these scales (e.g. \citealt{momi2008}). Finally, our calculations do not include any model for the feedback processes that accompany the formation of a young star (e.g. \citealt{mt08}). Given these uncertainties, the sink particles are unable to say exactly what the shape of the final IMF will be, but rather measure how the gas can fragment at a given scale (our resolution), and how these fragments are likely to evolve, assuming feedback processes play only a minor role over the timescales investigated in this study. These caveats should be borne in mind when interpreting the results in the following sections.

\section{The fragmentation of primordial gas}
\label{sec:frag}
In our comparison of the different simulations performed in this study, we will compare the properties of the clouds after roughly 10 percent of their mass has been accreted (although in some cases we will mention in passing what happens as the simulations are advanced further). In terms of looking at the ability of the clouds to fragment, comparing the different simulations at this point in their evolution ensures that they have turned the same faction of their initial number of Jeans masses into the sink particles. The exception in our analysis is the comparison between the 1000 \solmas Pop.\ III.1 and III.2 simulations, since in this case the Pop.\ III.2 clouds are initially more Jeans unstable. We now go on to describe the fragmentation in the Pop.\ III.1 and III.2 channels in some detail.

\subsection{Pop.\ III.1 clouds}
\label{pop31}
 The panels in Fig.~\ref{fig:image3.1} show the column density distribution in the Pop.\ III.1 simulations after 10 percent of their mass has been accreted onto the sink particles. The images show the inner 1300 AU of the cloud, centred on the first sink particle to form in each simulation. We can see that the clouds with $\Delta v_{\rm turb} \geq 0.2 c_{\rm s}$  fragment into small clusters of sink particles; the runs with  $\Delta v_{\rm turb} = 0.2 c_{\rm s}$, $0.4 c_{\rm s}$ and $0.8 c_{\rm s}$ form 5, 31 and 15 sink particles respectively. These calculations clearly demonstrate that turbulent subsonic motions are able to promote fragmentation in primordial gas clouds. Only in the case with $\Delta v_{\rm turb} = 0.1 c_{\rm s}$ does the cloud form just a single sink particle. In this case an extended disk builds up around the star, since the seed turbulence gives rise to some low level of rotation in the collapsing core,  but otherwise the gas contains little structure. 
 
Interestingly, there is no clear trend linking the number of fragments that form and the initial turbulent energy: the 0.4 \cs run fragments more than the 0.8 \cs run, despite containing only one quarter of the initial turbulent energy. The effects at play here are somewhat complex. First, the turbulent velocity field contained in the collapsing region will differ with each value of $\Delta v_{\rm turb}/c_{\rm s}$, since the different strengths of flow will push the gas around to different degrees. In addition, the nature of the turbulence that survives in the collapsing core will also affect the fragmentation. As we see from the images in Fig.~\ref{fig:image3.1}, the cloud with $\Delta v_{\rm turb} = 0.4 c_{\rm s}$ -- the most successful in terms of fragmentation -- forms a large disk-like structure. As we will discuss in \S \ref{sec:evol}, this configuration appears to aid the fragmentation of the infalling envelope. Therefore, much of the cloud's ability to fragment depends on the level of rotation that happens to become locked-up in the collapsing region. Further, and to a lesser degree, the extra delay in the collapse caused by the increased turbulent support also gives the cloud more time to wash out anisotropies in the gas (see Fig. \ref{fig:tsf} for the collapse times). Thus the ability of the cloud to fragment is a competition between these conflicting processes. For the randomly generated velocity field that we used in this study, the ability to fragment is better for $\Delta v_{\rm turb}/c_{\rm s} = 0.4$, than $\Delta v_{\rm turb}/c_{\rm s} = 0.8$, but we note that this may not always be the case. We stress that to make a quantitative statistical statement about the number of fragments that form as a function of the turbulent Mach number, we would need to run a series of different realizations of the turbulent velocity field in each case. Such a comparison lies outside the scope of our current study.  

The mass functions of the sink particles from the simulations that undergo fragmentation are shown in Fig.~\ref{fig:sinkmf}. For clarity, we have omitted the single 100 \solmas sink particle that forms in the 0.1 \cs cloud. For the 0.2 \cs and 0.8 \cs  clouds, we see that the sink masses cluster around some central value -- roughly 12 \solmas and 4 \solmas respectively -- while the 0.4 \cs cloud has a mass function that is skewed to lower masses, with a peak at around 1 \solmasp,  a sharp fall-off below this, and a broad distribution towards higher masses, extending up to around 13 \solmasp. 
We emphasize that the mass functions presented here (as well as in Fig. 10)
do not yet represent the final IMFs, as they correspond to an intermediate
time in the overall accretion process, where only 10\% of the cloud has been
accreted. They also might be affected by the numerical details of our
sink technique, in particular the absence of any sink-sink mergers, as discussed in Section~3.

The reason for the spread of masses becomes apparent when we look at the accretion properties of the sink particles, which are shown in Fig.~\ref{fig:maccp1}. Focusing on the 0.4 and 0.8 \cs  runs, and looking at the evolution of the individual sinks in more detail, we see that some appear to accrete rapidly and then suddenly stop. This behaviour is  typical of what is seen in simulations of bound, fragmenting
cores in the context of present-day star formation \citep{kb00,Klessen01,kb01,bvb04,sk04},  and is a result of velocity kicks from dynamical three (or more) body interactions. The mass accretion rate in such a  system \citep{bonnell01a} is given by
\begin{equation}
\label{equ:bh}
\dot{m_{*}} \propto \rho \frac{m_*^2}{v_{\rm rel}^3},
\end{equation}

\noindent where $m_{*}$ is the mass of the sink particle, $\rho$ is the gas density, and $v_{\rm rel}$ is the velocity of the sink particle relative to the gas. The ability of a sink to accrete more mass from the available reservoir is significantly reduced once its velocity increases. The effect is then exacerbated by the fact that an increase in velocity results in the sink particle moving to a more distant orbit (or even being kicked out of the  system entirely) and hence into a position where the gas density is lower. Since this sink particle is now in a position where further accretion is difficult, its siblings are able to accrete its `share' of the mass reservoir, with the majority going to those few sinks that sit right in the middle of the cloud's potential well. In general, as the sinks accrete from the background gas, they tend to move towards the centre, due to mass-loading. Further, their increased mass makes them more likely to survive dynamical encounters with 
 their less massive siblings. As such, the `rich get richer', with a few sinks ending up significantly more massive than the rest. The process is typically termed `competitive accretion', 
 \citep{bonnell97,bonnell01a,bonnell01b,bb06} and normally leads to the type of distribution of masses seen in our 0.4\cs run (Fig.~\ref{fig:sinkmf}).

Looking at the mass evolution of the individual sink particles, we also see that the formation of new sink particles occurs in bursts. The turbulence generates structure in the gas which is enhanced by the gravitational collapse as the gaseous envelope falls in towards the central system of sink particles.
The bursts in sink formation reflect the moments when these structures detach from the flow and become self-gravitating in their own right.

Figure~\ref{fig:maccp1} also shows the accretion rate of the cluster as a whole, and how that compares to an estimate made from the radial infall profile. To construct this estimate, we first compute the mass infall rate as a function of the radial distance $r$ from the densest SPH particle:
\begin{equation}
\label{equ:mdotinfall}
\dot{m}(r) = 4 \pi \,r^{2}\,\rho(r)\, v_{r}(r),
\end{equation}

\noindent where $\rho(r)$ is the gas density in a spherical shell with radius $r$ and width d$r$, and $v_{r}(r)$ is the radial velocity of the gas in this shell and all quantities are volume-averaged. Given the enclosed mass as a function of radial distance, $m_{\rm enc}(r)$, it is straightforward to convert from $\dot{m}(r)$ to an infall rate as a function of enclosed mass, $\dot{m}(m_{\rm enc})$, which we identify with the total mass accretion rate of the system of sink particles.
Figure~\ref{fig:maccp1} demonstrates that this estimate provides a fairly close match
to the actual accretion rate onto the sink particle population, except at very early times, which is a numerical artifact: the sink particles instantly accrete all gas within their accretion radius when they form.  At later times, the main difference between the estimated accretion rate and the true accretion rate is that the latter is significantly noisier, owing to the bursts of sink formation and the clumpiness of the infalling gas.  Finally, we note that the accretion rates are also fairly insensitive to the level of turbulence in the cloud, suggesting that random turbulent motions with the magnitudes considered here should not lead to a variation in the overall accretion rates of primordial stars or star clusters from minihalo to minihalo. Other sources of support against gravity, such as rotation and magnetic fields (assuming the latter can be efficiently generated, as in \citealt{tb04} and \citealt{schl10}),  are likely to play a greater role in 
regulating the accretion rate. Indeed, the important role played by rotation in regulating the infall and accretion of gas can be appreciated if we compare
the accretion rates measured in our simulations with those estimated or measured in previous studies
of Pop.\ III star formation starting from more realistic cosmological initial conditions \citep[e.g.][]{abn02,brlb04}. The absence of initial rotational support in our simulations leads to  higher infall velocities, and hence to an accretion rate that is a factor of a few larger than these previous values.

\subsection{Pop.\  III.2 clouds}
\subsubsection{Small clouds with 150 \solmas}
The first clouds we will examine in the Pop.\ III.2 case are those with an initial energy balance similar to those studied in the Pop.\ III.1 case: that is, clouds that have only a few Jeans masses initially. Since the Pop.\ III.2 channel is cooler at number densities around $10^{5}$ cm$^{-3}$, with a typical temperature of around 75K, the same initial number of Jeans masses requires that the clouds have a lower mass of 150 \solmasp. 

The column density distribution in these clouds after 10 percent of the gas has been accreted by the sink particles is shown in the column density images in Fig.~\ref{fig:image3.2}. We see from the images that these clouds undergo significantly less fragmentation than their Pop.\ III.1 counterparts, forming at most 3 sink particles after 10 percent of the cloud mass has been accreted. While these calculations demonstrate that this primordial star formation channel is susceptible to fragmentation if turbulence is present in the collapsing gas, it appears to be significantly more stable than the Pop.\ III.1 channel that gives rise to the first stars in the universe.

The main reason why this mode of star formation is less susceptible to fragmentation has to do with the thermal evolution of the gas as it collapses. In Fig.~\ref{fig:rhotevol} we show the temperature of the gas as a function of its number density, for the Pop.\ III.1 and Pop.\ III.2 cases, taken from the simulations with $\Delta v_{\rm turb} = 0.1\: {c_{\rm s}}$. The elevated $\mHt$ and HD fractions in the Pop.\ III.2 simulations do not provide enough cooling to keep the gas close to the CMB temperature at these
densities, owing to the increasing inefficiency of the HD as a coolant as it nears the critical density 
at which the populations of its rotational level reach their local thermodynamic equilibrium (LTE) values.
The gas therefore heats up significantly as it collapses, increasing its temperature from 75~K at
$n = 10^{5} \: {\rm cm^{-3}}$ to roughly 700~K at $n = 10^{7} \: {\rm cm^{-3}}$, corresponding to evolution
that is almost adiabatic. This sharp increase in temperature temporarily increases the local Jeans
mass at the center of the cloud, and significantly slows the collapse, giving the turbulence in the cloud
time to decay. In the absence of additional physical processes able to replenish the turbulence,
there is nothing to sustain the density inhomogeneities in the
cloud that act as the seeds for later fragmentation (since they are not yet self-gravitating), and so the
end result is a collapse with a much lower level of fragmentation than in most of the  Pop.\ III.1 simulations. A similar effect has previously been noted by \citet{to08} in their study of fragmentation
in very metal poor gas clouds. They find that for metallicities of around ${\rm Z} \sim 10^{-4.5} \:
{\rm Z_{\odot}}$, heat input due to three-body H$_{2}$ formation at $n \sim 10^{8} \: {\rm cm^{-3}}$
leads to a sharp jump in the gas temperature, which delays the collapse, reduces the elongation of
the collapsing core, and suppresses any fragmentation. Furthermore, \citet{yoh07} also briefly addressed this issue in their study of Pop.\ III.2 star formation, and showed that their simulated 
Pop.\ III.2 prestellar core would be stable against gravitational deformation at similar densities,
owing to its hard effective equation of state.

Although the evolution of the temperature with density is significantly different between the Pop.\ III.1 and Pop.\ III.2 channels, we see that the accretion rates are similar when we consider the inner 0.01 to 10 solar masses of the collapsing envelope (Fig.~\ref{fig:maccp2}),  consistent with the results from 
 \citet{yoh07}. However, if we consider the evolution of the accretion rates over the whole cloud, we see that at later times the two channels depart significantly from one another, with accretion occurring at
a significantly slower rate in the Pop.\ III.2 case (Fig.~\ref{fig:mdotcloud}). Also, at very early times, the accretion rates in the Pop.\ III.2 clouds are more sensitive to the level of turbulence than the Pop.\ III.1 clouds, and the turbulence has a somewhat stronger effect in delaying the onset of star formation in these calculations (Fig.~\ref{fig:tsf}). 

\subsubsection{Large clouds with 1000 \solmas}
Given that the Pop.\ III.2 clouds seem to be much more stable against fragmentation, it is worthwhile investigating whether they can be made to fragment when the gas is initially more Jeans unstable. Recent simulations of the formation of the first galaxies show that regions where Pop.\ III.2 star formation occurs are fed by cold, supersonic turbulent streams of gas \citep[e.g.][]{2008MNRAS.387.1021G,2008ApJ...685...40W}.
As such, the initial condition for the Pop.\ III.2 channel in this picture may have significantly more than one Jeans mass, due to the rapid assembly of the self-gravitating core. In this section we consider Pop.\ III.2 clouds that contain 1000 \solmas, but otherwise have the same properties as the 150 \solmas clouds (i.e., same initial temperature, density, and initial chemical composition). Since these clouds are colder than the Pop.\ III.1 clouds of the same mass and density, they are initially more Jeans unstable, having around 26 Jeans masses at the start of the simulation.

The results of two such calculations are shown in Fig.~\ref{fig:image3.2big}, in which the initial levels of turbulence have been set to 0.4 \cs and 0.8~$c_{\rm s}$. Again we show the evolution at the point where 10 percent of the cloud's mass has been converted into (or accreted onto) sink particles. Although these clouds are now initially more unstable than the Pop.\ III.1 clouds, they form significantly fewer sink particles, only 8 in the 0.4 \cs cloud and 7 in the 0.8 \cs cloud (see Fig.~\ref{fig:sinkmf32big}). The fact that the mass functions of the sink particles appear to have two peaks is due to the cloud undergoing two distinct bursts of sink formation.

The interesting result here is that these clouds exhibit {\em less} fragmentation than their Pop.\ III.1 counterparts, despite containing {\em more} Jeans masses in the initial configuration. The main reason for this is the gas in the Pop.\ III.2 clouds follows an extremely `stiff' effective equation of state (essentially adiabatic evolution), as can be seen from the temperature-density relationship shown in Fig.~\ref{fig:rhotevol}. Such a rapid increase in the temperature during the initial collapse makes generating structure in the gas difficult, helping to remove any anisotropies introduced by the turbulent flows. An additional effect is that for equations of state with effective adiabatic index of $\gamma \ga 4/3$ (as is the case with these clouds), the Jeans mass increases with increasing density. As such, the collapse halts until sufficient mass has been assembled, and the new Jeans mass has been reached. Since the cooling time is significantly longer than the free-fall time, the gas has a chance to remove structure that could potentially assist the fragmentation at higher densities, where the effective $\gamma$ is more similar to that found in the Pop.\ III.1 clouds. The combination of these effects results in a gas which is much more stable to fragmentation during collapse than in the Pop.\ III.1 simulations.

\section{Long term evolution of the infalling envelope}
\label{sec:evol}

More insight into the stability of the clouds against fragmentation can be gained from Fig.~\ref{fig:evol}, where we show the temperature and $\mHt$\, fraction as a function of density, for the 1000 \solmas Pop.\ III.1 and III.2 clouds in which the initial level of turbulence was 0.4~$c_{\rm s}$. The different colors correspond to different points in the evolution of the clouds, with red, green and blue corresponding to the formation of the first sink particle, 50 \solmas accreted, and 100 \solmas accreted, respectively. The temperature-density plots show the same behavior as reported in \citet{sgb10}. The particle evolution appears to diverge at a density of around $10^{10}$\,cm$^{-3}$, with one group of SPH particles heating up to a maximum temperature of around 7000K and the other staying around 1500K and cooling significantly at higher densities. As discussed in \citet{sgb10}, the hot part of the diagram corresponds to gas that falls in at a later stage in the evolution of the cloud. At that point, the enclosed gas mass is larger than at earlier times, and  so the free-fall velocity is correspondingly larger. The gas therefore shocks more strongly than at early times, causing it to become hot enough to collisionally dissociate its H$_{2}$ rapidly. With the H$_{2}$ gone, there is nothing to cool the gas until its temperature reaches $T \sim 7000 \: {\rm K}$. At this point, Lyman-$\alpha$ cooling becomes effective, allowing it to resist further heating. Although we see the same trends in both the Pop.\ III.1 and Pop.\ III.2 cases, in the latter the amount of gas departing from the `standard' temperature-density evolution is significantly reduced, since the elevated $\mHt$ fractions allow the post-shock gas to cool more effectively, limiting the temperature rise and allowing more of the gas to retain its H$_{2}$.

From the accompanying graphs we see that the $\mHt$ is rapidly dissociated as the temperature rises, but given the relatively low amounts of $\mHt$ present in the gas at these densities, the cooling provided by the dissociation is clearly unable to offset the compression. However we see that at higher densities the $\mHt$ suddenly reforms, coinciding with the region in the temperature-density diagrams where the gas is cold. This prompts the question of whether the cold gas is the result of an isochoric cooling instability, brought on by rapid $\mHt$ formation? Does the turbulence trigger an instability that naturally exists in pure primordial gas, but that is less effective in gas that has been influenced by the presence of previous star formation?

By harnessing the Lagrangian nature of SPH, the plots in Fig.~\ref{fig:ptrace} shed some light on this issue. They show the temporal evolution of the density, temperature and $\mHt$ fraction for several SPH particles that eventually become sink particles, and as such trace the conditions in the gas in the run-up to gravitational fragmentation and collapse. Again the simulations are those shown in Fig.~\ref{fig:evol}. The quantities are calculated by averaging over each particle's 50 nearest neighbors at each instant
in time. In the case of the Pop.\ III.1 cloud, these sinks are the last to form in the simulation, while in the Pop.\ III.2 cloud, the five lines represent all of the sink formation that occurs between the formation of the first and last sinks.

The figure shows a number of interesting features. First, we see that none of those SPH particles destined to become sinks undergo the rapid rise in temperature -- and accompanying loss of $\mHt$ -- that is shown in Fig.~\ref{fig:evol}. In contrast, their temperatures remain close to the temperatures found within the first collapsing core (see Fig.~\ref{fig:rhotevol}). This demonstrates that the cool particles seen in Fig.~\ref{fig:evol} do not come from regions that undergo shock heating and subsequent loss of $\mHt$. Instead, we see that they come from regions of gas that first undergo a relatively quiescent collapse, before being involved in several expansions and contractions. 

The fact that this first stage of the density evolution is fairly slow needs to be stressed: the free-fall times at densities of $n = 10^{9} \: {\rm cm^{-3}}$ and $n = 10^{10} \: {\rm cm^{-3}}$ are approximately 1600 yr and 500 yr, respectively. As such, these particles are not experiencing as much compression as those that end up losing their $\mHt$. In fact we see that they actually have a higher than average $\mHt$ fraction, when we compare them with the lower right-hand plot in Fig.~\ref{fig:evol}, a property that helps them remain fairly cool as they collapse. The reason why they experience less compression is that they are collapsing in a rotating structure that has been formed during the collapse of the turbulent gas.

The evolution of these particles and their immediate surroundings alters abruptly once they enter the dynamically complicated swirling regions that we can see in the column density figures. First, they collide with other material, which results in a sharp increase in their temperature and density. This increase in the density in turn increases the rate of $\mHt$ formation, which very rapidly turns them fully molecular. As they re-expand, the adiabatic cooling and the now significantly enhanced $\mHt$ line-cooling act together to reduce the temperature. In some cases this happens several times, but for all of these particles, the end result is the same and they find themselves in a clump of gas that is now Jeans unstable and fully molecular: the perfect conditions for forming a new protostar. Interestingly, we see similar behaviour leading up to the formation of the sink particles in both the Pop.\ III.1 and III.2 clouds. The relative lack of fragmentation in the Pop.\ III.2 case simply results from the lack of structure in the collapsing envelope, rather than any thermal properties of the gas at these high densities.

\section{Discussion}
\label{sec:chat}

The calculations presented in this study suggest that the first stars in the universe may form in small dense clusters, provided that the turbulent initial conditions we adopt are close to those found in minihalos. This suggests that the Population~III IMF covered a broad range in masses, possibly exhibiting a scale-free, power-law extension similar to the present-day case. Previous arguments in favor of a peaked IMF, in the shape of a narrow Gaussian or even a delta function, would then need to be revisited (e.g., Bromm \& Larson 2004). Although a similar prediction can be drawn from the study of Clark et al.~(2008, hereafter CGK08), there are some subtle differences. Our present study employs a fully self-consistent treatment of the thermodynamics, rather than the piece-wise polytropic equation of state (EOS) approach used by CGK08, which was taken as a fit to the detailed one-zone calculations of \citet{om05}. Although both start with similar (Pop.\ III.1) initial conditions, the fragmentation seen in the current simulations has a different origin. In the self-consistent treatment employed here, the fragmentation is driven by the complicated thermodynamics of high density clumpy gas as it enters the disk-like regions that surround the first protostar. In contrast, the fragmentation in CGK08 was evident at much lower densities, with the turbulent flows forming structures that were enhanced during the collapse, and rotation providing a `window of opportunity' for those structures to become gravitationally unstable in their own right, rather than being simply accreted onto the central protostellar core. Note however that the simulations in CGK08 also contained systematic, solid-body rotation in the initial conditions, which we do not study in this paper. This may help some of the structure to survive.
  
In general, the calculations presented in this paper suggest that the structure formed early on in the collapse is less likely to survive when the full thermodynamical behavior of the gas is taken into account. This is evident in the fact that the current calculations yield significantly fewer fragments than the Pop III simulations in CGK08 -- 25 sink particles for 19 \solmas of accreted gas in CGK08, compared to 31 sink particles for 100 \solmas of accreted gas in the current study. One feature that should be stressed (and which was originally pointed out in \citealt{sgb10}) is that the long-term thermodynamic evolution of the envelope differs significantly from the one-zone models. This brings into question the practice of using a piece-wise barotropic EOS as a proxy for the full thermodynamics in simulations that intend to study the evolution of the gas beyond the collapse of the first protostellar core.

Assuming that the turbulent initial conditions used in this paper are indeed representative of the gas in minihalos, it is worth considering the implications of the fragmentation that we see. The mass spectrum of the fragments in our calculations ranges from a few 0.1 \solmas to a few 10 \solmasp. We would not expect stellar feedback to change this result significantly, since previous studies \citep[e.g.][]{mt08} have shown that feedback becomes effective at limiting accretion onto Pop.\ III protostars only for protostellar masses greater than about 20 \solmasp. However, such feedback effects are expected to become important, and in particular to suppress further fragmentation (e.g., Krumholz et al. 2009) during the further evolution of the cloud. The fragments will then likely grow in mass, and possibly even merge with each other. It is very difficult to extrapolate to the final situation where accretion and merging stops. But our main result that the Population~III IMF was likely broad seems robust.  

As we have discussed above, it is difficult to reach any definitive
conclusions regarding the final Population~III masses and the resulting IMF.
It appears at least possible, however, that in
rare cases truly metal-free stars with masses less than 
$\sim 0.8$ \solmas could have formed, and would still be present in the Milky Way today, thus  providing a unique opportunity to directly probe the physical conditions at the end of the dark ages with Galactic observations. Current models of hierarchical galaxy formation predict that the first and most metal-poor halos to merge will become part of the bulge component of the resulting spiral galaxy \citep[e.g.][]{tum10}. It is therefore highly interesting to survey the bulge of the Milky Way for extremely metal-poor and metal-free stars. However, this is also very challenging as the bulge is far away, has high stellar density, and contains a wide range of stellar populations of all ages and metallicities. The odds of finding a few truly metal-free stars amongst millions of other stars are very low, or even zero due to pollution from the interstellar medium (Frebel et al. 2009).

With the same caveat as above, it is interesting to consider the implications
of a Pop.\ III IMF with a characteristic mass significantly smaller than
the canonical $100 \: {\rm M_{\odot}}$. For instance,
the flux of ionizing photons produced by a population of Pop.\ III  stars has a significant dependence
on the form of the Pop.\ III IMF, as does the metal-enrichment pattern produced by a collection of Pop.\ III supernovae \citep{tvs04}. Indeed, the abundance patterns observed to date in extremely metal-deficient stars in the Galactic halo \citep[see e.g.\ the review by][]{bc05} are far more consistent with an IMF that produces primarily core-collapse supernovae, with progenitor masses of 10--$40 \: {\rm M_{\odot}}$, rather than with an IMF that produces only very massive pair-instability supernovae, or PISNe \citep{jog10}. On the other hand, there may be subtle selection effects at work that bias current surveys against finding PISN-enriched stars (Karlsson et al. 2008). The basic argument here is that PISNe have such high metal yields that abundances in stars that form out of this material are already quite high (see Greif et al. 2010), and would therefore be missed in searches that target the lowest metallicities. The interpretation of the large carbon enhancements seen in 
 the population of carbon-enhanced metal-poor stars as the result of the enrichment of these stars by winds from binary companions that have passed through the AGB phase also implies the existence of a large number of intermediate-mass Pop.\ III stars, with masses $M = 1$--$8 \: {\rm M_{\odot}}$ \citep{tum07a,tum07b}. Again, there are alternative models to explain
the carbon enhancement in metal-poor stars in terms of nucleosynthesis
in faint supernovae (e.g., Iwamoto et al. 2005), in line with a higher
characteristic Pop.~III mass.

It is important to note that one factor that may limit the impact of turbulent fragmentation on the primordial IMF is if most Pop.\ III stars form in conditions resembling our Pop.\ III.2 clouds. It is relatively straightforward to show that most Pop.\ III stars will form in halos that have been affected in some fashion by a previous episode of Pop.\ III star formation \citep{ts09,ggbk10}. Thus, using the \citet{tm08} terminology, most Pop.\ III stars will be Pop.\ III.2 stars. However, it is less obvious how many of these stars will have formed out of gas that has been cooled to temperatures $T \ll 200 \: {\rm K}$ by HD. The effectiveness of HD cooling in these systems depends on the balance between the enhanced formation of H$_{2}$ and HD, owing to the enhanced initial fractional ionization in this gas, and the destruction of H$_{2}$ and HD due to Lyman-Werner band absorption of ultraviolet photons from an extragalactic background (Haiman, Abel \& Rees 2000; Johnson, Greif 
 \& Bromm 2008), or from local sources \citep{on99,gb01}. The relative importance of these effects has not been explored in great detail, and in any case, the outcome is likely to be sensitive to uncertainties in the microphysics that have only recently been resolved \citep{gsj06,ga08,kr10}.

To sum up, our calculations suggest that turbulent fragmentation may play an important role in
the formation of Pop.\ III stars, and may strongly influence the form of the Population III IMF.
However, the approximate nature of our calculation -- specifically, our simplified choice of initial conditions -- means that we should regard this at present as no more than a plausible hypothesis.
To better establish the role of  turbulent fragmentation in real Pop.\ III minihalos, we will need to
be able to simulate a representative sample of both Pop.\ III.1 and Pop.\ III.2 minihalos with 
sufficiently high resolution, such that they can resolve the turbulent flows in the gas at the moment when it becomes 
self-gravitating. This is a challenging prospect, lying far beyond of the scope of this preliminary
study, but our results suggest that it would prove extremely worthwhile.

\section{Summary}
\label{sec:sum}
We have explored the effects of subsonic turbulence on the gravitational collapse of primordial gas clouds. The study employed sink particles to model the run-away collapse of protostellar cores, which allowed us to follow the evolution of the collapsing clouds beyond the formation of the first protostar. The calculations also used a full time-dependent chemical network that accounts for the thermodynamic behaviour of the gas.  The current calculations contain neither magnetic fields nor feedback from the protostars. Our main findings can be summarized as follows:

\begin{itemize}

\item Turbulent primordial gas is unstable to fragmentation when one considers the evolution {\em beyond} the formation of the first protostellar core. 

\item Gas starting from conditions appropriate to Pop.\ III.1 collapse, rather than Pop.\ III.2 star formation, is more susceptible to fragmentation. As suggested by \citet{yoh07}, the thermal evolution of Pop.\ III.2 gas as it collapses helps to suppress further gravitational instability over and above the main collapse mode. However some fragmentation in the Pop.\ III.2 case is seen in this study, caused by the inhomogeneities introduced by the turbulence.

\item In the cases where fragmentation is efficient (in particular the Pop III.1 cloud with turbulent
rms velocity $\Delta v_{\rm turb} = 0.4 c_{\rm s}$), the masses of the fragments extend over a large range, which results in a distribution of stellar masses exhibiting a power-law extension towards high mass, as seen in the present-day IMF \citep{kroupa02, chabrier03}. However it should be noted that the exact form of the sink particle mass function may depend on our assumed initial conditions (see Appendix~\ref{ICdiscuss}) and the simplifications employed in the sink particle implementation itself.

\item Due to the relative lack of fragmentation in the Pop.\ III.2 clouds compared to the Pop.\ III.1 clouds, the stars formed in the Pop.\ III.2 calculations are of higher mass on average (for the same accreted mass) than their Pop.\ III.1 counterparts, even though the mass accretion rate of the star cluster as a whole is higher in the Pop.\ III.1 clouds.

\item Fragmentation tends to occur in gas which has temperatures of around 200 -- 400~K at densities above $10^{11} \: {\rm cm}^{-3}$, significantly lower than the 1000 -- 1500~K associated with the collapse of the first core. Such cold temperatures are a result of the expansion that occurs as gas enters the rotating, disk-like regions around the central core, coupled with relatively high fractions of $\mHt$, which can provide efficient line-cooling. These rotating structures are themselves a consequence of the angular momentum that is present in the turbulence.

\end{itemize}

In summary, we propose that if even small levels of turbulence, with velocity dispersions of order 20\% of the sound speed or more, are present in the baryonic component of dark matter minihalos, primordial stars are likely to be born in small stellar groups, rather than in isolation, and to have a wide range of stellar masses.  Further, the very first stars (Pop.\ III.1) may have lower masses on average than the second generation of stars (Pop.\ III.2), contrary to what has previously been assumed.

\begin{acknowledgments}

The authors would like to thank Tom Abel and Thomas Greif for stimulating discussions that helped shape this paper.
The work presented in this paper was assisted by the European Commission FP6 Marie Curie RTN CONSTELLATION (MRTN-CT-2006-035890). P.C.C.\ acknowledges support by the {\em Deutsche Forschungsgemeinschaft} (DFG) under grant KL 1358/5. R.S.K.\ acknowledges financial support
from the {\em Landesstiftung Baden-W\"urrtemberg} via their program International Collaboration II
(grant P-LS-SPII/18) and from the German {\em Bundesministerium f\"ur Bildung und Forschung} via
the ASTRONET project STAR FORMAT (grant 05A09VHA). R.S.K.\ furthermore acknowledges 
subsidies from the DFG under grants no.\ KL1358/1, KL1358/4, KL1358/5, KL1358/10, and KL1358/11,
as well as from a Frontier grant of Heidelberg University sponsored by the German Excellence Initiative.
R.S.K.  also thanks the Kavli Institute for Particle Astrophysics and Cosmology at Stanford University and
the Department of Astronomy and Astrophysics at the University of California at Santa Cruz  for their warm hospitality during a sabbatical stay in spring 2010.
V.B.\ acknowledges support from NSF grant AST-0708795 and NASA ATFP grant
NNX08AL43G. Part of the simulations were carried out at the Texas Advanced
Computing Center (TACC), under TeraGrid allocation TG-AST090003.
\end{acknowledgments}

\bibliographystyle{apj}

\appendix
\section{Chemistry}
\label{num:chem}
To model the chemical evolution of the metal-free gas, we use the network
for primordial hydrogen, helium and deuterium chemistry detailed in 
Table~\ref{tab:chem}. Our treatment of the hydrogen and helium chemistry 
largely follows \citet{ga08}, but our treatment of the deuterium chemistry is significantly
simplified. This simplification arises from our neglect of ${\rm D^{-}}$,
${\rm HD^{+}}$ and ${\rm D_{2}}$, none of which play a significant role
in controlling the HD abundance in the physical conditions relevant to our
present study. For the most part, our choice of rate coefficients also follows
\citet{ga08}.\footnote{Note that there are two typographical errors that we are
aware of in the set of reaction rate coefficients listed in Table A1 in \citet{ga08}.
First, the fitting function used to describe the rate coefficient for the reaction 
${\rm H_{2}} + {\rm H^{+}} \rightarrow {\rm H_{2}^{+}} + {\rm H}$ 
(reaction number 7 in their table) should use natural logarithms, and not base 10
logarithms, as listed. However, the listed fitting coefficients are correct. Second,
the temperature dependence of the reaction ${\rm H_{2}} + {\rm He} \rightarrow
{\rm He} + {\rm H} + {\rm H^{+}}$ (their reaction 24) should be $\exp(-35/T)$, 
and not $\exp(+35/T)$.}
We adopt case B rate coefficients for the recombination of H$^{+}$ and He$^{++}$, 
and treat He$^{+}$ recombination as decribed in Section 2.1.4 of \citet{ga08}. 
We adopt rate coefficients from \citet{gp98} for the associative detachment of 
H$^{-}$ ions by atomic hydrogen (reaction~2), and for
the mutual neutralization of H$^{-}$ by H$^{+}$ (reaction~5).
We also note that the uncertainties in these reaction rate coefficients discussed in 
\citet{gsj06} are unlikely to be significant in the high density, low ionization 
gas modelled in our present simulations.
For the rate coefficient of 
reaction 30, the three-body formation of H$_{2}$ with atomic hydrogen as
the third body
\begin{equation}
 {\rm H} + {\rm H} + {\rm H} \rightarrow {\rm H_{2}} + {\rm H},
\label{tbh2a}
\end{equation}
we adopt the rate coefficient proposed by \citet{g08}. The value of the rate
coefficient for this reaction is highly uncertain \citep{g08,tcg10}, but our 
chosen value lies intermediate between the fastest and slowest rates
to be found in the literature (\citealt{fh07} and \citealt{abn02}, respectively),
and is based on the best currently available data for the inverse reaction
(i.e.\ collisional dissociation of H$_{2}$ by H, reaction 9). The influence of the 
uncertainty in this rate coefficient has been studied in detail elsewhere 
\citep{tcg10}. To fix the rate coefficient for the analogous reaction
\begin{equation}
 {\rm H} + {\rm H} + {\rm H_{2}} \rightarrow {\rm H_{2}} + {\rm H_{2}},
\end{equation}
we follow \citet{pss83} and assume that the rate coefficient is one-eighth
of the size of the rate coefficient adopted for reaction~\ref{tbh2a}. For
three-body formation in which neutral helium is the third body, we follow
\citet{ga08} and use a rate coefficient originally taken from \citet{wk75}. 
We assume, following \citet{fh07}, that the rate coefficients for the three-body
formation of HD (reactions 43--45) are the same as those for the corresponding
H$_{2}$ formation reactions (nos.\ 30--32).

We adopt values for the rate coefficients for collisional dissociation of H$_{2}$ 
by H, H$_{2}$ and He that are consistent with our choices for the three-body 
association rates in the sense that each pair of rate coefficients individually satisfies 
\begin{equation}
\frac{k_{\rm form}}{k_{\rm dest}} = K,
\end{equation}
where $k_{\rm form}$ is the rate coefficient for three-body association, 
$k_{\rm dest}$ is the rate coefficient for collisional dissociation, and where
the equilibrium constant $K$ is given in all three cases by \citep{fh07}
\begin{equation}
K = 1.05 \times 10^{-22} T^{-0.515} \exp \left(\frac{52000}{T} \right).
\end{equation}
This procedure is necessary in order to ensure the correct chemical behaviour
of the gas at high densities and temperatures, where the H$_{2}$ formation 
and destruction timescales are both short, and most of the gas is close to 
chemical equilibrium.

\begin{deluxetable}{llc}
\tablecaption{Reactions included in our chemical model \label{tab:chem}}
\tablewidth{0pt}
\tablehead{
\colhead{No.\ } & \colhead{Reaction} & \colhead{References} }
\startdata
1 & $\mH + \me  \rightarrow  \Hm + \gamma $ & 1 \\
2 & $\Hm  + \mH  \rightarrow \mHt + \me$ & 2 \\
3 & $\mH + \Hp  \rightarrow  \mHtp + \gamma $ & 3 \\
4 & $\mH + \mHtp \rightarrow \mHt + \Hp$ & 4 \\
5 & $\Hm  + \Hp  \rightarrow \mH + \mH$ & 2 \\
6 & $\mHtp + \me  \rightarrow  \mH + \mH $ & 5 \\
7 & $\mHt + \Hp  \rightarrow  \mHtp + \mH $ & 6 \\
8 & $\mHt + \me  \rightarrow  \mH + \mH +  \me$ & 7 \\
9 & $\mHt + \mH  \rightarrow  \mH + \mH + \mH$  & 8 \\
10 & $\mHt + \mHt \rightarrow  \mHt + \mH + \mH$ & 9, 10 \\ 
11 & $\mHt + \He \rightarrow \mH + \mH + \He$ & 11 \\
12 &  $\mH + \me  \rightarrow  \Hp + \me + \me $ & 12 \\
13 & $\Hp  + \me  \rightarrow \mH +  \gamma$ & 13 \\
14  &  $\Hm + \me  \rightarrow  \mH + \me + \me $ & 12 \\
15  &  $\Hm + \mH  \rightarrow  \mH + \mH + \me $ & 12 \\
16  &  $\Hp + \Hm  \rightarrow  \mHtp + \me $ & 14 \\
17 &  $\He + \me  \rightarrow  \Hep + \me + \me $ & 12 \\
18 & $\Hep + \me \rightarrow \Hepp + \me + \me$ &  12 \\
19 & $\Hep + \me \rightarrow \He + \gamma$ & 15, 16 \\
20 & $\Hepp + \me \rightarrow \Hep + \gamma$ & 13 \\
21 & $\Hm + \mHtp \rightarrow \mHt + \mH$ & 17 \\
22 & $\Hm + \mHtp \rightarrow \mH + \mH + \mH$ & 17 \\
23 & $\mHt + \me \rightarrow \Hm + \mH$ & 18 \\
24 & $\mHt + \Hep \rightarrow \He + \mH + \Hp$ & 19 \\
25 & $\mHt + \Hep \rightarrow \mHtp + \He$ & 19\\
26 & $\Hep + \mH \rightarrow \He + \Hp$ & 20 \\
27 & $\He + \Hp \rightarrow \Hep + \mH$ & 21 \\
28 & $\Hep + \Hm \rightarrow \He + \mH$ & 22 \\
29 & $\He + \Hm \rightarrow \He + \mH + \me$ & 23 \\
30 & $\mH + \mH + \mH \rightarrow \mHt + \mH$ & 24 \\
31 & $\mH + \mH + \mHt \rightarrow \mHt + \mHt$ & 25 \\
32 & $\mH + \mH + \He \rightarrow \mHt + \He$ & 26 \\
33 & $\Dp + \me \rightarrow \mD + \gamma$ & 27 \\
34  &  $\mD + \Hp \rightarrow  \mH + \Dp $ & 28 \\
35 &  $\mH + \Dp  \rightarrow  \mD + \Hp $ & 28 \\
36 & $\mHt + \mD \rightarrow \hd + \mH$ & 29 \\
37 & $\mHt + \Dp \rightarrow \hd + \Hp$ & 30 \\
38 &  $\hd + \mH  \rightarrow  \mHt + \mD $ & 31 \\
39 & $\hd + \Hp \rightarrow \mHt + \Dp$ & 30 \\
40 & $\mD + \me \rightarrow \Dp + \me + \me$ & 27 \\
41 & $\Hep + \mD \rightarrow \Dp + \He$ & 32 \\
42 & $\He + \Dp \rightarrow \mD + \Hep$ & 32 \\
43 & $\mD + \mH + \mH \rightarrow \hd + \mH$ & See text \\
44 & $\mD + \mH + \mHt \rightarrow \hd + \mHt$ & See text \\
45 & $\mD + \mH + \He \rightarrow \hd + \He$ & See text \\ 
\enddata
\tablecomments{
1: \citet{WIS79},  2: \citet{gp98}, 3: \citet{RAM76}, 4: \citet{KAR79},
5: \citet{SCH94}, 6: \citet{SAV04}, 7: \citet{tt02a}, 8: \citet{MSM96},  
9: \citet{MAR98}, 10: \citet{sk87}, 11: \citet{drcm87}, 12: \citet{JAN87}, 
13: \citet{FER92}, 14: \citet{POU78}, 15: \citet{hs98}, 16: \citet{ap73}, 
17: \citet{dl87}, 18: \citet{sa67}, 19: \citet{b84}, 20: \citet{z89}, 
21: \citet{kldd93}, 22: \citet{ph94}, 23: \citet{h82}, 24: \citet{g08}, 
25: \citet{g08}, rescaled as in \citet{fh07}, 26: \citet{wk75},
27: Same as corresponding H reaction, 28: \citet{sav02},  
29: Fit to data from \citet{mie03}, 30: \citet{ger82}, 31: \citet{s59}, 
32: \citet{ga08}}
\end{deluxetable}

\section{Cooling function and thermodynamics}
\label{sec:chem}
The dominant coolant in our simulations is molecular hydrogen. To model
rotational and vibrational line emission from H$_{2}$, we use the detailed
cooling function described in \citet{ga08}, that includes contributions 
from collisions of H$_{2}$ with H, He, H$_{2}$, protons and electrons. At
densities $n \simgreat 10^{9} \: {\rm cm^{-3}}$, the strongest of the H$_{2}$ 
lines become optically thick, reducing its effectiveness as a coolant. To model 
H$_{2}$ cooling in this regime, we use an approach based on the Sobolev 
approximation that was first used in models of primordial star formation 
by \citet{yoha06}. We write the H$_{2}$ cooling rate in optically thick gas as
\begin{equation}
\Lambda_{\rm H_{2}} = \sum_{\rm u,l} \Delta E_{\rm ul} A_{\rm ul} \beta_{\rm esc, ul} 
n_{\rm u},
\end{equation}
where $n_{\rm u}$ is the number density of hydrogen molecules in upper energy
level $u$, $\Delta E_{\rm ul}$ is the energy difference between this upper level
and a lower level $l$, $A_{\rm ul}$ is the spontaneous radiative transition rate
for transitions between $u$ and $l$, and $\beta_{\rm esc, ul}$ is the escape
probability associated with this transition, i.e.\ the probability that the
emitted photon can escape from the region of interest. We take values for 
the level energies from the compilation made available by P.~G.~Martin on
his website\footnote{http://www.cita.utoronto.ca/$\sim$pgmartin/h2.html}
and for the radiative transition rates
from \citet{wsd98}, and we fix $n_{\rm u}$ by assuming that the 
H$_{2}$ level populations are in LTE. The problem of modelling optically thick
H$_{2}$ cooling thereby reduces to one of computing the escape probability for
each transition. We follow \citet{yoha06} and write the escape probability for
the transition ${\rm u \rightarrow l}$ as
\begin{equation}
\beta_{\rm esc, ul} = \frac{1 - \exp(-\tau_{\rm ul})}{\tau_{\rm ul}},
\end{equation}
where we approximate $\tau_{\rm ul}$ as 
\begin{equation}
\tau_{\rm ul} = \alpha_{\rm ul} L_{\rm s} \label{tau}
\end{equation}
where $\alpha_{\rm ul}$ is the line absorption coefficient and $L_{\rm s}$ is the
Soloblev length. In the classical, one-dimensional spherically symmetric case,
the Sobolev length is given by
\begin{equation}
L_{\rm s} = \frac{v_{\rm th}}{|{\rm d}v_{\rm r} / {\rm d}r|},
\end{equation}
where $v_{\rm th}$ is the thermal velocity, and ${\rm d}v_{\rm r} / {\rm d}r$ is the
radial velocity gradient. In our inherently three-dimensional flow, we generalize 
this as
\begin{equation}
L_{\rm s} = \frac{v_{\rm th}}{|\nabla \cdot {\mathbf v}|},
\end{equation}
following \citet{nk93}. If the velocity dispersion of the gas is very small,
then $L_{\rm s}$ can become very large, much larger than the size of the
collapsing core. To ensure that we do not reduce the H$_{2}$ cooling rate in
this case to an artificially low value, we take as our actual length scale in
Equation~\ref{tau} the smallest of the Soloblev length and the local Jeans 
length, $L_{\rm J}$.

Since the line absorption coefficient $\alpha_{\rm ul}$ is linearly proportional
to the number density of H$_{2}$, we can write $\tau_{\rm ul}$ as 
\begin{equation}
\tau_{\rm ul} = \left(\frac{\alpha_{\rm ul}}{n_{\rm H_{2}}} \right) N_{\rm H_{2}, eff},
\end{equation}
where $N_{\rm H_{2}, eff} \equiv n_{\rm H_{2}} L_{\rm s}$ is an effective H$_{2}$
column density, and where $\alpha_{\rm ul} / n_{\rm H_{2}}$ is a function only of
temperature. We therefore tabulate the cooling rate per H$_{2}$ molecule in
the optically thick limit as a function of two parameters: the gas temperature
$T$ and the effective H$_{2}$ column density $N_{\rm H_{2}, eff}$, and compute 
cooling rates during the simulations by interpolation from a pre-generated
look-up table.

At densities $n > 10^{14} \: {\rm cm^{-3}}$, a second form of H$_{2}$ cooling becomes 
important, called collision-induced emission. Although H$_{2}$ molecules have no 
electric dipole, the interacting pair in a collision of H$_{2}$ with H, He or H$_{2}$
briefly acts as a `supermolecule' with an non-zero electric dipole, and a 
hence a non-zero probability of emitting or absorbing a photon through a 
dipole transition. Because the collision time is very short, the resulting
collision-induced transition lines are very broad, effectively merging 
into a continuum \citep[for more details, see e.g.][]{fromm93}. We include
this process in our cooling function, using a rate taken from \citet{ra04}.
We crudely account for the reduction of the CIE cooling rate by continuum
absorption at very high number densities using the following prescription 
(M. Turk, private communication)
\begin{equation}
\Lambda_{\rm CIE, thick} = \Lambda_{\rm CIE, thin} \times {\rm min}\left(\frac{1 - 
e^{-\tau_{\rm CIE}}}{\tau_{\rm CIE}}, 1 \right)
\end{equation}
where 
\begin{equation}
\tau_{\rm CIE} = \left(\frac{n}{1.4 \times 10^{16} \: {\rm cm^{-3}}} \right)^{2.8}.
\end{equation}
However, we note that optical depth effects do not strongly affect the CIE
cooling rate for densities below our threshold for sink particle creation
(see \S\ref{sec:ics}), and so the approximate nature of this opacity cutoff is unlikely
to significantly affect our results. 

In addition to H$_{2}$ ro-vibrational line emission and CIE cooling, our
cooling function contains a number of other radiative processes: electronic
excitation of H, He and He$^{+}$, cooling from the recombination of H$^{+}$
and He$^{+}$, Compton cooling and bremsstrahlung. Details of our treatment of 
these processes can be found in \citet{gj07}.

We also account for changes in the thermal energy of the gas due to changes
in its chemical makeup. Specifically, we include the effects of 
cooling due to the collisional ionization of H, He and He$^{+}$, and 
due to the destruction of H$_{2}$ by charge transfer and by collisional 
dissociation, as well as heating due to the three-body H$_{2}$ formation. 
The balance between cooling due to H$_{2}$ collisional dissociation and 
heating due to three-body H$_{2}$ formation plays a very important role in
regulating the thermal evolution of the gas at densities 
$n \simgreat 10^{8} \: {\rm cm^{-3}}$. 

\section{Evolution of the Bonnor-Ebert spheres}
\label{ICdiscuss}

\begin{figure}[t]
	\centerline{		
    		\includegraphics[height=3.4in]{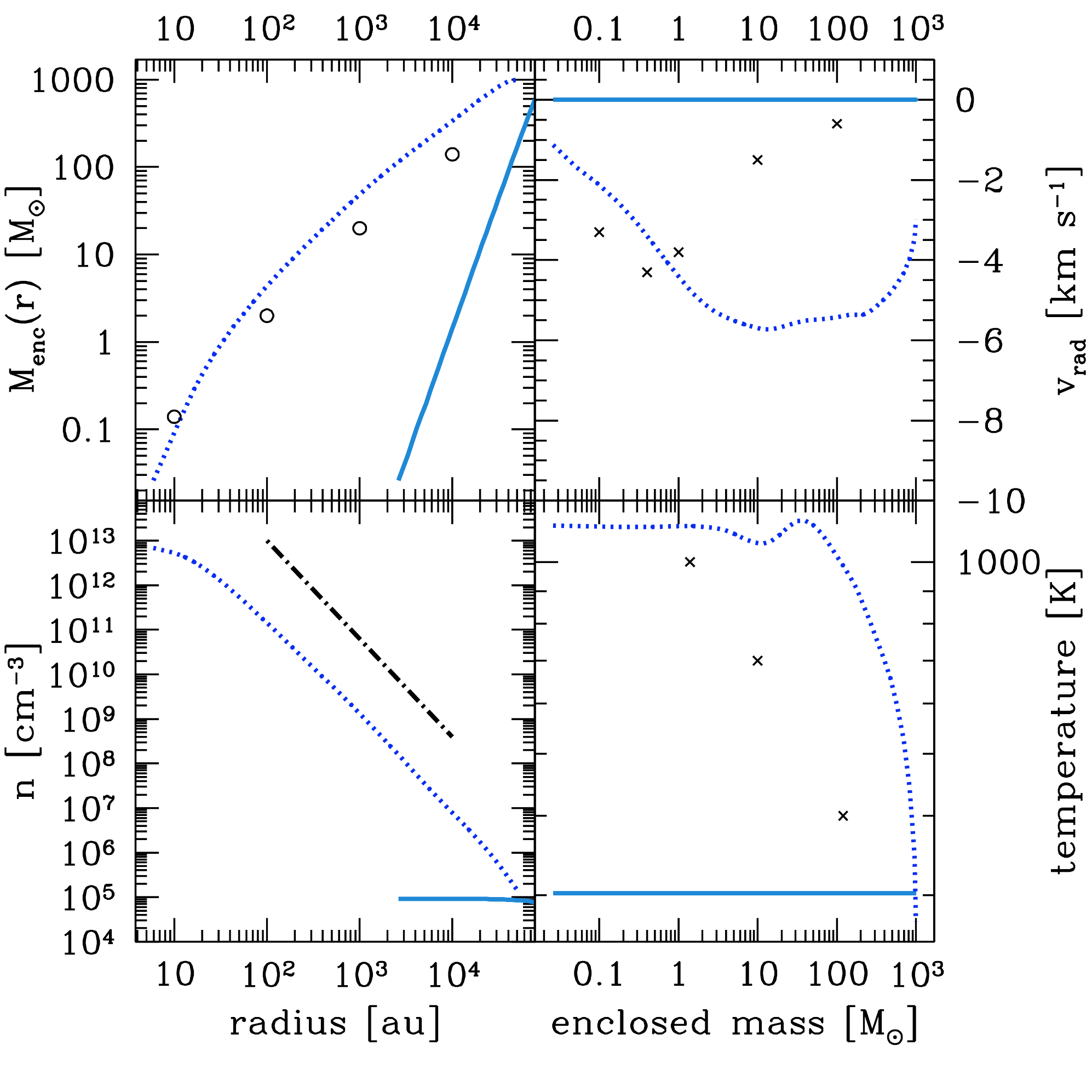}	
		\includegraphics[height=3.4in]{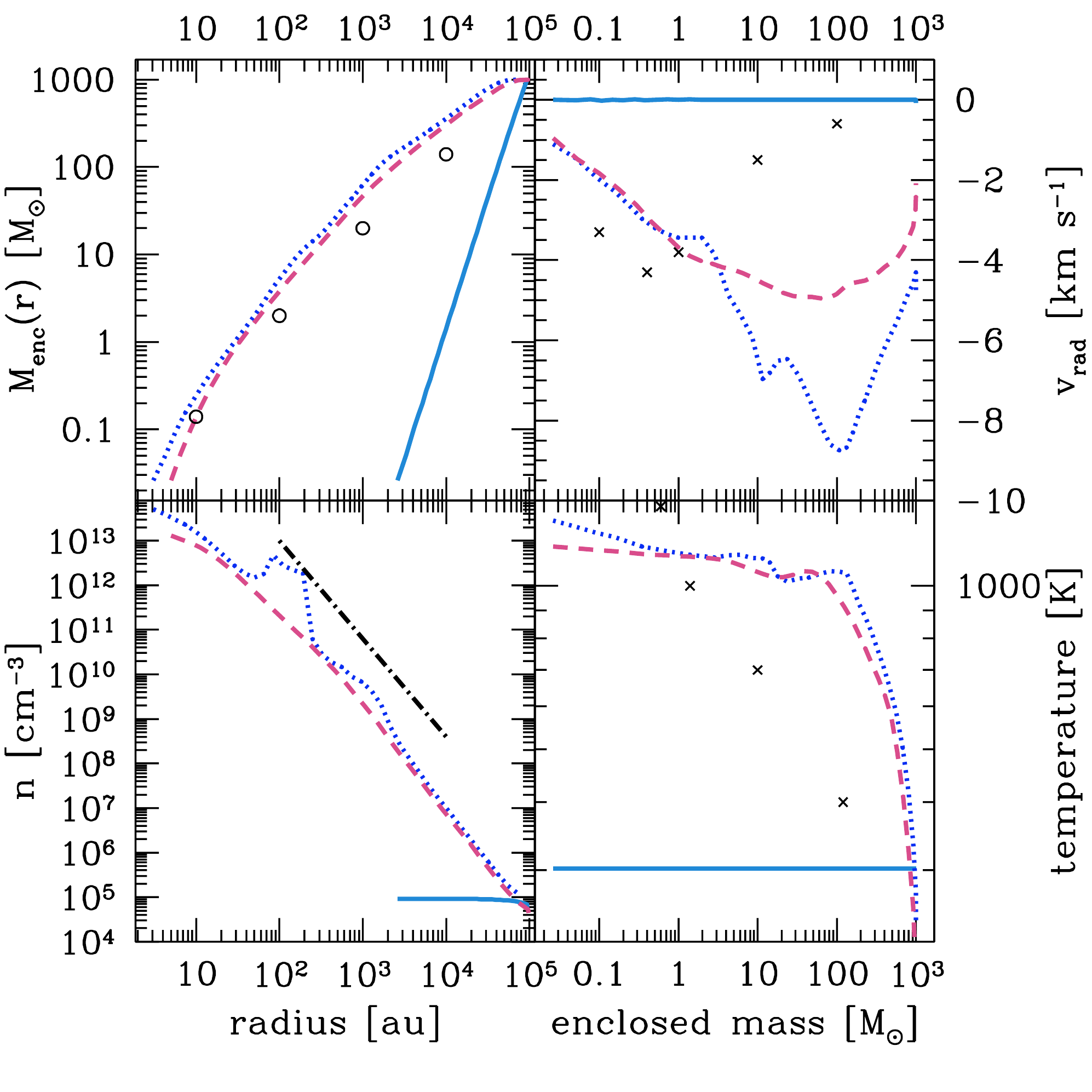}
		}
\caption{\label{fig:profiles} Radial profiles from the Pop III.1 clouds with $\Delta v_{\rm turb} = 0.1 c_{\rm s}$ (left panel) and $\Delta v_{\rm turb} = 0.4 c_{\rm s}$ (right panel). The lines show the initial conditions (solid line), the state of the gas just before the creation of the first sink (dotted line). In the case of the $\Delta v_{\rm turb} = 0.4 c_{\rm s}$ cloud, we also show the results of a collapse that has a factor of 4 less external pressure (dashed line). The dot-dashed line shows a slope of $n \propto r^{-2.2}$. The crosses and circular points denote approximate values taken, respectively, from \citet{yoha06} and \citet{abn02}}
\end{figure}

Given our somewhat arbitrary choice of initial conditions, it is prudent to ask whether they are applicable to primordial star formation, and in particular, whether the choice of the initial density profile affects the details of the collapse. In Fig.~\ref{fig:profiles}, we show, for the Pop III.1 clouds with 0.1 and 0.4 $c_{\rm s}$ turbulence, the number density and enclosed mass as a function of radius and how the radial velocity and temperature varies as a function of the enclosed mass. In the case of the 0.1$c_{\rm s}$ cloud, the radial profiles (mass and number density) are much like those presented for the standard Pop III studies \citep{abn02, yoha06}, in which fully cosmological initial conditions were used. In particular, by the time that the first sink particle forms in this run, the gas has developed the same $n \propto r^{-2.2}$ radial density profile as found in the fully cosmological runs, demonstrating that the density profile at this stage is insensitive to the shape of the initial density profile. 

We do however see differences between the radial velocity and temperature profiles, compared to those published in the literature. The infall velocities at this stage are somewhat higher than those seen in fully cosmological studies, but it should be noted that we do not include systemic rotation in our study, and so a major source of support is missing. However, aside from the higher values of the radial velocity, the shape of the profile is also different. In the runs presented in this study, the infall velocities of the shells enclosing 1 \solmas and greater are systematically higher than those seen in the cosmological simulations published to date, with a peak in the radial velocity profile at about 10-100 \solmas (depending on the level of turbulent support) instead of around 0.3 \solmasp. Some of this difference in the radial velocity profiles is reflected in the somewhat steeper than reported temperature profiles, since the gas will reacts to the increased $p {\rm d} V$ heating. However the temperature profiles are also affected by the differences in the chemistry, and in particular, our choice of the three-body H$_{2}$ formation rate. 

So the obvious question is, do our high infall velocities make fragmentation more likely than it would be reality? To address this, we performed another Pop III.1 simulation with 0.4 $c_{\rm s}$ turbulence, but this time we set the confining external pressure to a factor of 4 less than was used in the original simulations. The radial and enclosed mass profiles at the point just before the first sink particle is formed are shown as the dashed lines in Fig. \ref{fig:profiles} (right-hand panel). We see that radial velocity profile is affected by the change in boundary pressure, resulting in a factor of 4 reduction in the infall speed. Compared to the original run, which formed 31 sink particles, this simulation forms 22 sinks when 10 percent of the gas has been accreted. While this suggests that the fragmentation we see is sensitive to the details of the collapse, and we are likely overestimating the number of fragments formed (at least at the scale of our sink particles) when one compares to the standard Pop III collapse profiles, it seems that this is not the main cause of the fragmentation. The gas still fragments, and the processes leading up to the fragmentation are the same as those discussed in \S \ref{sec:evol}.

Finally, we note that there is also new evidence to suggest that the radial collapse profiles seen in the literature so far, may not be representative of all Pop III star formation. For example, while the radial density profiles from the studies by \citet{abn02} and \citet{yoha06} are similar to the newer, higher-resolution, cosmological simulations of \citet{tao09} and \citet{tcg10}, the radial (or enclosed mass) profiles of the other quantities in the newer calculations show considerable variation. In particular, the velocity and temperature profiles in \citet{tcg10} lie somewhere in between those found by \citet{yoha06} and our 0.4 $c_{\rm s}$ turbulent cloud. A very high-resolution study of the onset of the initial gravitational instability in the baryonic component of dark matter minihalos, may help to establish whether these differences between the studies of, say, \citet{abn02} and  \citealt{tao09} are due to the increased resolution and improved physical treatment used in the latter study (e.g.\ the inclusion of the effects of three-body H$_2$ formation heating), or just reflect the effects of cosmic variance.

\end{document}